\documentclass[11pt]{article}

\usepackage{mathtools}
\usepackage{sn-preamble}
\usepackage[margin=1in]{geometry}

\newcommand{\rnote}[1]{\footnote{\color{purple}\textbf{Rocco:} {#1}}}
\newcommand{\snote}[1]{\footnote{\color{blue}\textbf{Shivam:} {#1}}}

\newenvironment{proofof}[1]{\par{\noindent \textit{Proof of #1}.}}{\qed\par}

\newcommand{\VarInf}{\mathbf{VarInf}}
\newcommand{\GInf}{\Inf^{\calG}}
\newcommand{\rinn}{r_{\mathsf{in}}}
\newcommand{\ignore}[1]{}

\def\colorful{1}

\ifnum\colorful=1

\newcommand{\red}[1]{{\color{red} {#1}}}

\newcommand{\gray}[1]{{\color{gray}{#1}}}

\fi
\ifnum\colorful=0

\newcommand{\red}[1]{{{#1}}}

\newcommand{\gray}[1]{{{#1}}}
\fi

\title{Convex Influences}
\ignore{\rnote{is this too flashy/imprecise a title? I do think it will be convenient for us to be able to use the phrase ``convex influence'' to refer to this influence notion throughout the paper}}

\author{
Anindya De\\
\small{\sl University of Pennsylvania} \and Shivam
Nadimpalli  \\ \small{\sl Columbia University} \and Rocco A.
Servedio \\ \small{\sl Columbia University}
}

\date{\small{\today}}

\begin{document}

\maketitle

\begin{abstract}
	We introduce a new notion of influence for symmetric convex sets over Gaussian space, which we term ``convex influence''.\ignore{a notion of the \emph{influence of direction $v$ on $K$}, where $K \subset \R^n$ is a symmetric convex set and $v \in \mathbb{S}^{n-1}$ is a unit vector.}  We show that this new notion of  influence\ignore{for symmetric convex sets in $n$-dimensional Gaussian space} shares many of the familiar properties of influences of variables for monotone Boolean functions $f: \bn \to \bits.$  
	
	Our main results for convex influences give Gaussian space analogues of many important results on influences for monotone Boolean functions. These include (robust) characterizations of extremal functions, the Poincar\'{e} inequality,  the Kahn-Kalai-Linial theorem \cite{KKL:88}, a sharp threshold theorem of Kalai \cite{Kalai:04}, a stability version of the Kruskal-Katona theorem due to O'Donnell and Wimmer \cite{OWimmer:09}, and some partial results towards a Gaussian space analogue of Friedgut's junta theorem \cite{Friedgut:98}. The proofs of our results for convex influences use very different techniques than the analogous proofs for Boolean influences over $\bn$.
	Taken as a whole, our results extend the emerging analogy between symmetric convex sets in Gaussian space and monotone Boolean functions from $\bn$ to $\bits.$
\end{abstract}

\thispagestyle{empty}



\newpage

\setcounter{page}{1}


\section{Introduction} \label{sec:intro}

\noindent {\bf Background: An intriguing analogy.} This paper is motivated by an intriguing, but at this point only partially understood, analogy between \emph{monotone Boolean functions over the hypercube} and \emph{symmetric convex sets in Gaussian space.} Perhaps the simplest manifestation of this analogy is the following pair of easy observations:  since a Boolean function $f: \bn \to \bits$ is monotone if $f(x) \leq f(y)$ whenever $x_i \leq y_i$ for all $i$, it is clear that ``moving an input up towards $1^n$'' by flipping bits from $-1$ to $1$ can never decrease the value of $f$.  Similarly, we may view a symmetric\footnote{A set $K \subseteq \R^n$ is symmetric if $-x \in K$ whenever $x \in K.$} convex set $K \subseteq \R^n$ as a 0/1 valued function, and it is clear from symmetry and convexity that  ``moving an input in towards the origin'' can never decrease the value of the function. 

The analogy extends far beyond these easy observations to involve many analytic and algorithmic aspects of monotone Boolean functions over $\bn$ under the uniform distribution and symmetric convex subsets of $\R^n$ under the Gaussian measure.  Below we  survey some known points of correspondence  (several of which were only recently established) between the two settings:

\begin{enumerate}

\item {\bf Density increments.} The well-known Kruskal-Katona theorem \cite{Kruskal:63,katona1968theorem} gives quantitative information about how rapidly a monotone $f: \bn \to \bits$ increases on average as the input to $f$ is ``moved up towards $1^n$.''  Let $f: \bn \to \zo$ be a monotone function and let $\mu_f(j)$ be the fraction of the ${n \choose j}$ many weight-$j$ inputs for which $f$ outputs 1; the Kruskal-Katona theorem implies (see e.g.~\cite{lovasz2007combinatorial}) that if $k = cn$ for some $c$ bounded away from 0 and 1 and $\mu_f(k) \in [0.1,0.9]$, then $\mu_f(k+1) \ge \mu_f(k)  + \Theta(1/n).$ Analogous ``density increment'' results for symmetric convex sets are known to hold in various forms, where the analogue of moving an input in $\bn$ up towards $1^n$ is now moving an input in $\R^n$ in towards the origin, and the analogue of $\mu_f(j)$ is now $\alpha_r(K)$, which is defined to be the fraction of the origin-centered radius-$r$ sphere $r\mathbb{S}^{n-1}$ that lies in $K$.  For example, Theorem~2 of the recent work \cite{DS21-weak-learning} shows that if $K \subseteq \R^n$ is a symmetric convex set (which we view as a function $K: \R^n \to \zo$) and $r=\Theta(\sqrt{n})$ satisfies $\alpha_r(K) \in [0.1,0.9]$, then $\alpha_K(r(1-1/n) ) \geq \alpha_K(r) + \Theta(1/n)$. 

\item {\bf Weak learning from random examples.} Building on the above-described density increment for symmetric convex sets, \cite{DS21-weak-learning}  showed that any symmetric convex set can be learned to accuracy $1/2 + \Omega(1)/\sqrt{n}$ in $\poly(n)$ time given $\poly(n)$ many random examples drawn from ${\cal N}(0,1)^n$. \cite{DS21-weak-learning} also shows that any $\poly(n)$-time weak learning algorithm (even if allowed to make membership queries) can achieve accuracy no better than $1/2 + O(\log(n)/\sqrt{n})$. These results are closely analogous to the known (matching) upper and lower bounds for $\poly(n)$-time weak learning of monotone functions with respect to the uniform distribution over $\bn$:  Blum et al.~\cite{BBL:98} showed that $1/2 + \Theta(\log(n)/\sqrt{n})$ is the best possible accuracy for a $\poly(n)$-time weak learner (even if membership queries are allowed), and O'Donnell and Wimmer \cite{OWimmer:09} gave a $\poly(n)$ time weak learner that achieves this accuracy using random examples only.

\item {\bf Analytic structure and strong learning from random examples.} \cite{BshoutyTamon:96} showed that the Fourier spectrum of any $n$-variable monotone Boolean function over $\bn$ is concentrated in the first $O(\sqrt{n})$ levels. 
Analogously, \cite{KOS:08} showed that the same concentration holds for the first $O(\sqrt{n})$ levels of the Hermite spectrum\footnote{The Hermite polynomials form an orthonormal basis for the space of square-integrable real-valued functions over Gaussian space; the Hermite spectrum of a function over Gaussian space is analogous to the familiar Fourier spectrum of a function over the Boolean hypercube. See~\Cref{sec:prelims} for details.} of the indicator function of any convex set.  
In both cases this concentration gives rise to a learning algorithm, using random examples only, running in $n^{O(\sqrt{n}
)}$ time and learning the relevant class (either monotone Boolean functions over the $n$-dimensional hypercube or  convex sets under Gaussian space) to any constant accuracy.

\item {\bf Qualitative correlation inequalities.} The well-known Harris-Kleitman theorem  \cite{harris60,kleitman66} states that monotone Boolean functions are non-negatively correlated: any monotone $f,g: \bn \to \zo$ must satisfy $\E[fg] - \E[f]\E[g] \geq 0$.  The Gaussian Correlation Inequality \cite{roy14} gives an exactly analogous statement for symmetric convex sets in Gaussian space: if $K,L \subseteq \R^n$ are any two symmetric convex sets, then $\E[K L] - \E[K]\E[L] \geq 0$, where now expectations are with respect to ${\cal N}(0,1)^n$.

\item {\bf Quantitative correlation inequalities.}  Talagrand  \cite{Talagrand:96} proved the following \emph{quantitative} version of the Harris--Kleitman inequality: for monotone $f,g: \bn \to \zo$,
\begin{equation}~\label{eq:Talagrand}
\E[f g] - \E[f] \E[g] \ge \frac{1}{C} \cdot \Psi\left(\sum_{i=1}^n \Inf_i[f] \Inf_i(g)\right).
\end{equation} 
Here $\Psi(x) := x/\log(e/x)$, $C>0$ is an absolute constant, $\Inf_i[f]$ is the influence of coordinate $i$ on $f$ (see \Cref{sec:prelims}), and the expectations are with respect to the uniform distribution over $\bn$.
In a recent work \cite{DNS20} proved a closely analogous quantitative version of the Gaussian Correlation Inequality: for $K, L$ symmetric convex subsets of $\R^n$, 
\begin{equation}
\label{eq:DNS}
\E[KL] - \E[K]\E[L] \geq {\frac 1 C} \cdot \Upsilon\left(\sum_{i=1}^n \widetilde{K}(2e_i)\widetilde{L}(2e_i)\right),
\end{equation}
where $\Upsilon: [0,1] \to [0,1]$ is $\Upsilon(x) = \min\left\{x,\frac{x}{\log^2\left(1/x\right)}\right\}$, $C>0$ is a universal constant, $\widetilde{K}(2e_i)$ denotes the degree-2 Hermite coefficient in direction $e_i$ (see~\Cref{sec:prelims}), and expectations are with respect to ${\cal N}(0,1)^n$.

\end{enumerate}

We remark that in many of the above cases the proofs of the two analogous results (Boolean versus Gaussian) are very different from each other even though the statements are quite similar.\ignore{ For example, the original Kruskal-Katona theorem has a combinatorial proof whereas the Gaussian analogue described in the first list item above can be proved using either elementary geometric and probabilistic arguments (see \cite{DS21-weak-learning}) or spherical isoperimetry (see \cite{DS21-weak-learning}).} For example, the Harris-Kleitman theorem has a simple one-paragraph proof by induction on $n$, whereas the Gaussian Correlation Inequality was a famous conjecture for four decades before Thomas Royen proved it in 2014.

\medskip

\noindent {\bf Motivation.}
We feel that the examples presented above motivate a deeper understanding of this ``Boolean/Gaussian analogy.''  This analogy may be useful in a number of ways; in particular, via this connection known results in one setting may suggest new questions and results for the other setting.\footnote{Indeed, the recent Gaussian density increment and weak learning results of \cite{DS21-weak-learning}  were inspired by the Kruskal-Katona theorem and the weak learning algorithms and lower bounds of \cite{BBL:98} for monotone Boolean functions. Similarly, the recent quantitative version of the Gaussian Correlation Theorem established in \cite{DNS20} was motivated by the existence of Talagrand's quantitative correlation inequality for monotone Boolean functions.}
Thus the overarching goal of this paper is to strengthen the analogy between monotone Boolean functions over $\bn$ and symmetric convex sets in Gaussian space.   We do this through the study of a new notion of \emph{influence} for symmetric convex sets in Gaussian space.  

\subsection{This Work:  A New Notion of Influence for Symmetric Convex Sets}

Before presenting our new notion of influence for symmetric convex sets in Gaussian space, we first briefly recall the usual notion for Boolean functions.  For $f: \bn \to \bn$, the \emph{influence of coordinate $i$ on $f$}, denoted $\Inf_i[f]$, is $\Pr[f(\bx) \neq f(\bx^{\oplus i})]$, where $\bx$ is uniform random over $\bn$ and $\bx^{\oplus i}$ denotes $\bx$ with its $i$-th coordinate flipped.  It is a well-known fact (see e.g.~Proposition~2.21 of \cite{ODbook}) that for monotone Boolean functions $f$, we have $\Inf_i[f] = \widehat{f}(i)$, the degree-1 Fourier coefficient corresponding to coordinate $i$.

Inspired by the relation $\Inf_i[f] = \widehat{f}(i)$ for influence of monotone Boolean functions, and by the close resemblance between \Cref{eq:Talagrand} and \Cref{eq:DNS}, \cite{DNS20} proposed
to define the \emph{influence of $K$ along direction $v$,} for $K: \R^n \to \zo$ a symmetric convex set and $v \in \mathbb{S}^{n-1}$, to be
	\[ \Inf_v[K] := -\widetilde{K}(2v),\]
	the (negated) degree-2 Hermite coefficient\footnote{We observe that if $K$ is a symmetric set then since its indicator function is even, the degree-1 Hermite coefficient $\widetilde{K}(v)$ must be 0 for any direction $v$.} of $K$ in direction $v$ (see \Cref{def:csc-influence} for a detailed definition). \cite{DNS20} proved that this quantity is non-negative for any direction $v$ and any symmetric convex $K$ (see \Cref{prop:influence-nonneg}). They also defined the \emph{total influence of $K$} to be
\begin{equation} \label{eq:TInf}
\TInf[f] := \sum_{i=1}^n \Inf_{e_i}[f]
\end{equation}
and observed that this definition is invariant under different choices of orthonormal basis other than $e_1,\dots,e_n$, 
but did not explore these definitions further.
	
The main contribution of the present work is to carry out an in-depth study of this new notion of influence for symmetric convex sets. For conciseness, and to differentiate it from other influence notions (which we discuss later), we will sometimes refer to this new notion as ``convex influence.'' 

Inspired by well known results about influence of monotone Boolean functions, we establish a number of different results about convex influence which show that this notion shares many properties with the familiar Boolean influence notion.  Intriguingly, and similar to the Boolean/Gaussian analogy elements discussed earlier, while the statements we prove about convex influence are quite closely analogous to known results about Boolean influences, the proofs and tools that we use (Gaussian isoperimetry, Brascamp-Lieb type inequalities, theorems from the geometry of Gaussian space such as the $S$-inequality~\cite{s-inequality}, etc.) are very different from the ingredients that underlie the corresponding results about Boolean influence. 

\subsection{Results and Organization}

We give an overview of our main results below.

\paragraph{Basics, examples, Margulis-Russo, and extremal functions.} We begin in \Cref{sec:basics} by working through some basic properties of our new influence notion.  After analyzing some simple examples in \Cref{subsec:influence-examples}, we next show in \Cref{subsec:russo-margulis} that the total convex influence for a symmetric convex set is equal to (a scaled version of) the rate of change of the Gaussian volume of the set as the variance of the underlying Gaussian is changed. This gives an alternate characterization of total convex influence, and may be viewed as an analogue of the Margulis-Russo formula for our new influence notion.
We continue in \Cref{subsec:extremal} by giving some straightforward characterizations of extremal symmetric convex sets vis-a-vis our influence notion, namely the ones that have the largest individual influence in a single direction and the largest total influence. As one would expect, these extremal functions are the Gaussian space analogues of the Boolean dictator and majority functions respectively.   Next, we compare our new influence notion with some other previously studied notions of influence over Gaussian space (\Cref{subsec:other-notions-influence}).
These include the ``geometric influences'' that were studied by \cite{keller2012geometric} as well as the standard notion (from the analysis of functions over product probability domains, see e.g.~Chapter~8 of \cite{ODbook}) of the expected variance of the function along one coordinate when all other coordinates are held fixed.

\paragraph{Total influence lower bounds.}
In \Cref{sec:poincare-kkl} we give two lower bounds on the total convex influence (\Cref{eq:TInf}) for symmetric convex sets, which are closely analogous to the classical Poincar\'{e} and KKL Theorems. Our KKL analogue is quadratically weaker than the KKL theorem for Boolean functions; we conjecture that a stronger bound in fact holds, which would quantitatively align with the Boolean variant (see Item~1 of \Cref{sec:discussion}). Our proofs, which are based on the ``$S$-inequality'' of Lata\l a and Oleskiewicz \cite{s-inequality} and on the Gaussian isoperimetric inequality, are quite different from the proofs of the analogous statements for Boolean functions.

\paragraph{(A consequence of) Friedgut's junta theorem.} In \Cref{sec:our-friedgut} we establish a convex influences analogue of a consequence of Friedgut's junta theorem. Friedgut's junta theorem states that any Boolean $f: \bn \to \zo$ with small total influence must be close to a junta. This implies that for any monotone $f: \bn \to \zo$ with small total influence, ``averaging out'' over a small well-chosen set of input variables (the variables on which the approximating junta depends) results in a low-variance function. We prove a closely analogous statement for symmetric convex sets with small total convex influence, thus capturing a convex influence analogue of this consequence of Friedgut's junta theorem. (We conjecture that a convex influence analogue holds for Friedgut's original junta theorem; see Item~2 of \Cref{sec:discussion}.)

\paragraph{Sharp thresholds for functions with all small influences.} In \Cref{sec:sharp-threshold} we establish a ``sharp threshold'' result for symmetric convex sets in Gaussian space, which is analogous to a sharp threshold result for monotone Boolean functions due to Kalai \cite{Kalai:04}. Building on earlier work of Friedgut and Kalai \cite{FriedgutKalai:96}, Kalai \cite{Kalai:04} showed that if $f: \bn \to \zo$ is a monotone Boolean function and $p \in (0,1)$ is such that (i) all the $p$-biased influences of $f$ are $o_n(1)$ and (ii) the expectation of $f$ under the $p$-biased measure is $\Theta(1)$, then $f$ must have a ``sharp threshold'' in the following sense: the expectation of $f$ under the $p_1$-biased measure ($p_2$-biased measure, respectively) is $o_n(1)$ ($1-o_n(1)$, respectively) for some $p_1 < p < p_2$ with $p_2 - p_1 = o_n(1)$.  For our sharp threshold result, we prove an analogous statement for symmetric convex sets, where now ${\cal N}(0,\sigma^2)$ takes the place of the $p$-biased distribution over $\bn$ and the $\sigma$-biased convex influences (see \Cref{def:sigma-biased-influence}) take the place of the $p$-biased influences.  Interestingly, the sharpness of our threshold is quantitatively better than the known analogous result \cite{Kalai:04} for monotone Boolean functions; see \Cref{sec:sharp-threshold} for an elaboration of this point.  
\ignore{Our second sharp threshold result is an analogue of the celebrated result of Friedgut and Kalai \cite{FriedgutKalai:96}. This result states that if a monotone $f: \bn \to \zo$ satisfies condition (ii) above and moreover has ``sufficient symmetry,'' then it must have a sharp threshold (with a stronger quantitative bound than is established in \cite{Kalai:04}, thanks to the symmetry assumption).  We similarly show (as an easy consequence of our first sharp threshold result described above)that if a symmetric convex set has $\sigma$-biased Gaussian measure bounded away from 0 and 1 and has ``sufficient symmetry,'' then it must have a sharp threshold. 
}
 
\paragraph{A stable density increment result.} Finally, in \Cref{sec:robust-kk}, we use our new influence notion to give a Gaussian space analogue of a ``stability'' version of the Kruskal-Katona theorem due to O'Donnell and Wimmer \cite{OWimmer:09}.  In \cite{OWimmer:09} it is shown that the $\Omega(1/n)$ density increment of the Kruskal-Katona theorem (see Item~1 at the beginning of this introduction)  can be strengthened to $\Omega(\log(n)/n)$ as long as a ``low individual influences''-type condition holds.  We analogously show that a similar strengthening of the Gaussian space density increment result mentioned in Item~1 earlier can be achieved under the condition that the convex influence in every direction is low.
 
 \subsection{Techniques}
 
We give a high-level overview here of the techniques for just one of our results, namely our analogue of the KKL theorem, \Cref{thm:our-kkl}.  Several of our other results either employ similar tools (for example, our robust density increment result, \Cref{thm:KK-robust-convex}, and our main sharp threshold result, \Cref{thm:threshold-Friedgut-Kalai}) or else build off of \Cref{thm:our-kkl} (for example, our analogue of a consequence of Friedgut's junta theorem, \Cref{thm:weak-friedgut-convex}).

The KKL theorem states that if $f: \bn \to \bits$ has every coordinate influence small, specifically $\max_{i \in [n]} \Inf_i[f] \leq \delta$, then the total influence of $f$ must be large compared to $f$'s variance, specifically it must hold that $\TInf[f] = \Omega(\Var[f] \log(1/\delta))$.  This is a dramatic strengthening of the Poincar\'{e} inequality (which only states that $\TInf[f] \geq \Var[f]$) and is a signature result in the analysis of Boolean functions with many applications.  The classical proof of the KKL theorem is based on hypercontractivity \cite{Bon70,Bec75}, and only recently \cite{EldanGross:20,KKKMS:21} have proofs been given which avoid the use of hypercontractivity.

Our convex influences analogue of the KKL theorem states that if $K$ is a symmetric convex set and the convex influence $\Inf_v[K]$ in every direction $v \in \mathbb{S}^{n-1}$ is at most $\delta$ and $\delta \leq \Var[K]/10$, then the total convex influence $\TInf[K]$ must be at least $\Omega\pbra{\Var[K]\sqrt{\log\pbra{\frac{\Var[K]}{\delta}}}}.$ 
Our proof does not employ hypercontractivity but instead uses tools from convex geometry. It proceeds in two main conceptual steps:

\begin{enumerate}

\item First, we use a Brascamp--Lieb-type inequality due to Vempala \cite{Vempalafocs10} to argue that the maximum convex influence of $K$ in any coordinate can be lower bounded in terms of the Gaussian volume of $K$ and its ``width'' (equivalently, the radius of the largest origin-centered ball contained in $K$, which is called the \emph{in-radius} of $K$ and is denoted $\rinn(K)$).  This lets us show that
$\rinn(K) \geq \Omega(\sqrt{\ln(\Var[K]/\delta)})$ (see \Cref{eq:kkl-second-goal}).

\item Next, we argue that $\TInf[K] \geq {\frac 1 {\sqrt{\pi}}}\Var\sbra{K}\cdot\rinn(K)$ (see \Cref{eq:kkl-first-goal}), which together with the lower bound on $\rinn(K)$ gives the result.
 This is shown using our Margulis-Russo analogue, the Gaussian Isoperimetric Theorem, and concavity of the Gaussian isoperimetric function.

\end{enumerate}

\subsection{Discussion and Future Work} \label{sec:discussion}

We believe that much more remains to be discovered about this new notion of influences for symmetric convex sets.  We list some natural concrete (and not so concrete) questions for future work: 

\begin{enumerate}

\item {\bf A stronger KKL-type theorem for convex influences?} We conjecture that the factor of $\sqrt{\log(\Var[K]/\delta)}$ in our KKL analogue, \Cref{thm:our-kkl}, can be strengthened to $\log(1/\delta)$.  As witnessed by \Cref{eg:solid-cube}, this would be essentially the strongest possible quantitative result, and would align closely with the original KKL theorem \cite{KKL:88}.

\item {\bf An analogue of Friedgut's theorem for convex influences?}
As described earlier, our \Cref{thm:weak-friedgut-convex} establishes a Gaussian space analogue of a consequence of Friedgut's Junta Theorem \cite{Friedgut:98} for Boolean functions over $\bn$.  The following would give a full-fledged Gaussian space analogue of Friedgut's Junta Theorem:

\begin{conjecture} [Friedgut's Junta Theorem for convex influences] \label{conj:Friedgut}
Let $K \subseteq \R^n$ be a convex symmetric set with $\Inf[K] \leq I$.  Then there are $J \leq 2^{O(I/\eps)}$ orthonormal directions $v^1,\dots,v^{J} \in \mathbb{S}^{n-1}$ and a symmetric convex set $L \subseteq \R^n$, such that

\begin{enumerate}

\item  $L(x)$ depends only on the values of $v^1 \cdot x, \dots,v^{J} \cdot x$, and
\item $\Pr_{\bx \sim \calN(0, 1)^n}[K(\bx) \neq L(\bx)] \leq \eps.$

\end{enumerate}

\end{conjecture}

\item {\bf Are low-influence directions (almost) irrelevant?}  Related to the previous question, we note that it seems to be surprisingly difficult to show that low-influence directions ``don't matter much'' for convex sets.  For example, it is an open question to establish the following, which would give a dimension-free robust version of the last assertion of \Cref{prop:influence-nonneg}:

\begin{conjecture} \label{conj:low-inf}
Let $K \subseteq \R^n$ be symmetric and convex, and suppose that $v \in \mathbb{S}^{n-1}$ is such that $\Inf_v[K] \leq \eps.$ Then there is a symmetric convex set $L$ such that 

\begin{enumerate}

\item $L(x)$ depends only on the projection of $x$ onto the $(n-1)$ dimensional subspace orthogonal to $v$, and
\item $\Pr_{\bx \sim \calN(0, 1)^n}[K(\bx) \neq L(\bx)] \leq \tau(\eps)$ for some function $\tau$ depending only on $\eps$ (in particular, independent of $n$) and going to 0 as $\eps \to 0.$

\end{enumerate}

\end{conjecture}

While the corresponding Boolean statement is very easy to establish, natural approaches to \Cref{conj:low-inf} lead to open (and seemingly challenging) questions regarding dimension-free stable versions of the Ehrhard-Borell inequality   \cite{Figalli:20,Zvavitch:20}.

\item {\bf Algorithmic results?} Finally, a broader goal is to further explore the similarities and differences between the theory of convex symmetric sets in Gaussian space and the theory of monotone Boolean functions over $\bn$.  One topic where the gap in our understanding\ignore{ of monotone Boolean functions and (symmetric) convex sets} is particularly wide is the algorithmic problem of \emph{property testing}.   The problem of testing monotonicity of functions from $\bn$ to $\bits$ is rather well understood, with the current state of the art being an $\tilde{O}(n^{1/2})$-query upper bound and an $\tilde{\Omega}(n^{1/3})$-query lower bound \cite{KMS15,CWX17}. In contrast, the problem of testing whether an unknown region in $\R^n$ is convex (with respect to the standard normal distribution) is essentially wide open, with the best known upper bound being $n^{O(\sqrt{n})}$ queries \cite{chen2017sample} and no nontrivial lower bounds known.

\end{enumerate}

\section{Preliminaries} \label{sec:prelims}

In this section we give preliminaries setting notation and recalling useful background on convex geometry, log-concave functions, and Hermite analysis over $\calN\pbra{0, \sigma^2}^n$.

\subsection{Convex Geometry and Log-Concavity}
\label{subsec:log-concave}

Below we briefly recall some notation, terminology and background from convex geometry and log-concavity. Some of our main results employ relatively sophisticated results from these areas; we will recall these as necessary in the relevant sections and here record only basic facts.\ignore{ (for example, the $S$-inequality of Lata\l a and Oleskiewicz \cite{s-inequality} is crucial to the main results of \Cref{sec:poincare-kkl}),}  For a general and extensive resource we refer the interested reader to \cite{aga-book}. 

We identify sets $K \sse \R^n$ with their indicator functions $K : \R^n \to \zo$, and we say that $K\sse\R^n$ is \emph{symmetric} if $K(x) = K(-x)$. We write $B_r$ to denote the origin-centered ball of radius $r$ in $\R^n$.  If $K \subseteq \R^n$ is a nonempty symmetric convex set then we let $\rinn(K)$ denote $\sup_{r \geq 0} \{r: B_r \subseteq K\}$ and we refer to this as the \emph{in-radius} of $K$. 
\ignore{We record the following easy consequence of the separating hyperplane theorem:\rnote{I think we are likely to use this somewhere, right? Should we note (as LO do) that this is half the width of $K$ (should we define the width of $K$)?}

\begin{fact} \label{fact:slab}
Let $K \subseteq \R^n$ be a symmetric convex set with in-radius $\rinn < \infty.$ Then there is a direction $v \in \mathbb{S}^{n-1}$ such that $K \subseteq \{x \in \R^n: |v \cdot x| \leq \rinn\}.$
\end{fact}
}

Recall that a function $f : \R^n \to \R_{\geq 0}$ is \emph{log-concave} if its domain is a convex set and it satisfies $f(\theta x + (1-\theta)y) \geq f(x)^\theta f(y)^{1-\theta}$ for all $x, y \in \mathrm{domain}(f)$ and $\theta \in [0,1]$. In particular, the $0/1$-indicator functions of convex sets are log-concave. 

Recall that the \emph{marginal} of  $f: \R^n \to \R$ on the set of variables $\{i_1,\dots,i_k\}$ is obtained by integrating out the other variables, i.e. it is the function
\[
g(x_{i_1},\dots,x_{i_k}) = \int_{\R^{n-k}} f(x_1,\dots,x_n) dx_{j_1} \dots dx_{j_{n-k}},
\]
where $\{j_{1},\dots,j_{n-k}\} = [n] \setminus \{i_1,\dots,i_k\}$.  We recall the following fact:
\begin{fact}[\cite{Dinghas,Leindler,Prekopa, Prekopa2} (see Theorem 5.1, \cite{LV07})] 
\label{fact:marginal-log-concave}
	All marginals of a log-concave function are log-concave.
\end{fact}
The next fact follows easily from the definition of log-concavity:
\begin{fact} [\cite{Ibragimov:56}, see e.g.~\cite{An:95}]
\label{fact:unimodal-log-concave}
	A one-dimensional log-concave function is unimodal. 
\end{fact}

\subsection{Gaussian Random Variables}

We write $\bz \sim \calN(0,1)$ to mean that $\bz$ is a standard Gaussian random variable, and will use the notation 
\[\varphi(z) := \frac{1}{\sqrt{2\pi}}e^{-x^2/2} \qquad\text{and}\qquad \Phi(z) := \int_{-\infty}^z \varphi(t)\,dt\] 
to denote the pdf and the cdf of this random variable.

Recall that a non-negative random variable $\br^2$  is distributed according to the chi-squared distribution $\chi^2(n)$ if $\br^2 = \bg_1^2 + \cdots + \bg_n^2$ where $\bg \sim {\cal N}(0,1)^n,$ and that a draw from the chi distribution $\chi(n)$ is obtained by making a draw from $\chi^2(n)$ and then taking the square root.  

We define the \emph{shell-density function for $K$}, $\alpha_K : [0,\infty) \rightarrow [0,1]$, to be
\begin{equation} \label{eq:sdf}
\alpha_K(r) := \Prx_{\bx \in r\mathbb{S}^{n-1}} [\bx \in K],
\end{equation}
where the probability is with respect to the normalized Haar measure over $r\mathbb{S}^{n-1}$;
so $\alpha_K(r)$ equals the fraction of the origin-centered radius-$r$ sphere which lies in $K.$ We observe that if $K$ is convex and symmetric then $\alpha_K(\cdot)$ is a nonincreasing function.
 A view which will be sometimes useful later is that $\alpha_K(r)$ is the probability that a random Gaussian-distributed point $\bg \sim N(0,1)^n$ lies in $K$, conditioned on $\|\bg\|=r.$ 

\ignore{
We recall the following tail bound:\rnote{Check that we use this and the next fact, delete if we don't}
\begin{lemma} [Tail bound for the chi-squared distribution \cite{Johnstone01}] \label{lem:johnstone}
Let $\br^2 \sim \chi^2(n)$.
Then we have
\[\Prx\big[|\br^2-n| \geq tn\big] \leq e^{-(3/16)nt^2}\quad\text{for all $t \in [0, 1/2)$.}\]
It follows that  for $\br \sim \chi(n)$, 
\[
\Prx \big[ \sqrt{{n}/{2}} \le \br \le \sqrt{{3n}/{2}} \big] \ge 1-  e^{-\frac{3n}{64}}. 
\]
\end{lemma}
The following fact about the anti-concentration of the chi distribution will be useful:
\begin{fact} \label{fact:chi-squared-2}
For $n > 1$, the maximum value of the pdf of the chi distribution $\chi(n)$ is at most $1$, and hence for any interval $I=[a,b]$ we have 
$\Pr_{\br^2 \sim \chi^2(n)}[\br \in [a,b]] \leq b-a.$
\end{fact}
}

\subsection{Hermite Analysis over $\calN\pbra{0, \sigma^2}^n$}

Our notation and terminology here follow Chapter~11 of \cite{ODbook}. We say that an $n$-dimensional \emph{multi-index} is a tuple $\alpha \in \N^n$, and we define 
\begin{equation} \label{eq:index-notation}
|\alpha| := \sum_{i=1}^n \alpha_i.
\end{equation}

We write $\calN(0, \sigma^2)^n$ to denote the $n$-dimensional Gaussian distribution with mean $0$ and variance $\sigma^2$, and denote the corresponding measure by $\gamma_{n, \sigma}(\cdot)$. When the dimension $n$ is clear from context we simply write $\gamma_\sigma(\cdot)$ instead, and sometimes when $\sigma=1$ we simply write $\gamma$ for $\gamma_{1}$. For $n \in \N_{> 0}$ and $\sigma > 0$, we write $L^2\pbra{\R^n, \gamma_{\sigma}}$ to denote the space of functions $f: \R^n \to \R$ that have finite second moment $\|f\|_2^2$ under the Gaussian measure $\gamma_\sigma$, that is: 
\[
\|f\|_2^2 = \Ex_{\bz \sim \calN\pbra{0, \sigma^2}^n} \left[f(\bz)^2\right]^{1/2} < \infty.
\]
We view $L^2\pbra{\R^n, \gamma_\sigma}$ as an inner product space with $\la f, g \ra := \E_{\bz \sim \calN\pbra{0,\sigma^2}^n}[f(\bz)g(\bz)]$ for $f, g \in L^2\pbra{\R^n, \gamma_\sigma}$. 
We define ``biased Hermite polynomials,'' which yield an orthonormal basis for $L^2\pbra{\R^n, \gamma_{\sigma}}$:

\begin{definition}[Hermite basis]
	For $\sigma > 0$, the \emph{$\sigma$-biased Hermite polynomials} $(h_{j,\sigma})_{j\in\N}$ are the univariate polynomials defined as
	$$
	h_{j,\sigma}(x) := h_j\pbra{\frac{x}{\sigma}},
	\quad\text{where}\quad h_j(x) := \frac{(-1)^j}{\sqrt{j!}} \exp\left(\frac{x^2}{2}\right) \cdot \frac{d^j}{d x^j} \exp\left(-\frac{x^2}{2}\right).$$
\end{definition}

\begin{fact} [Easy extension of Proposition~11.33, \cite{ODbook}] \label{fact:biased-hermite-orthonormality}
	For $n \geq 1$ and $\sigma > 0$, the collection of $n$-variate $\sigma$-biased Hermite polynomials given by $(h_{\alpha, \sigma})_{\alpha\in\N^n}$ where
	$$h_{\alpha, \sigma}(x) := \prod_{i=1}^n h_{\alpha_i, \sigma}(x)$$
	forms a complete, orthonormal basis for $L^2(\R^n, \gamma_{\sigma})$. 
	\end{fact}

Given a function $f \in L^2(\R^n, \gamma_\sigma)$ and $\alpha \in \N^n$, we define its \emph{($\sigma$-biased) Hermite coefficient on} $\alpha$ to be $\widetilde{f}_\sigma(\alpha) := \la f, h_{\alpha, \sigma} \ra$. It follows that $f$ is uniquely expressible as $f = \sum_{\alpha\in\N^n} \widetilde{f}_\sigma(\alpha)h_{\alpha, \sigma}$ with the equality holding in $L^2(\R^n, \gamma_\sigma)$; we will refer to this expansion as the \emph{($\sigma$-biased) Hermite expansion} of $f$. When $\sigma = 1$, we will simply write $\wt{f}(\alpha)$ instead of $\wt{f}_\sigma(\alpha)$ and $h_\alpha$ instead of $h_{\alpha, 1}$.
Parseval's and Plancharel's identities hold in this setting: 

\begin{fact} \label{fact:hermite-plancharel}
	For $f, g \in L^2(\R^n, \gamma_\sigma)$, we have:
\begin{align*}\la f, g\ra &= \Ex_{\bz\sim\calN(0,\sigma^2)^n}[f(\bz)g(\bz)] = \sum_{\alpha\in \N^n}\widetilde{f}_\sigma(\alpha)\widetilde{g}_\sigma(\alpha), \tag{Plancherel}\\
\la f, f\ra& = \Ex_{\bz\sim\calN(0,\sigma^2)^n}[f(\bz)^2] = \sum_{\alpha\in \N^n}\widetilde{f}_\sigma(\alpha)^2. \tag{Parseval}
\end{align*}
\end{fact}

The following notation will sometimes come in handy. 

\begin{definition} \label{def:hermite-coeff-along-direction}
	Let $v \in \mathbb{S}^{n-1}$ and $f \in L^2\pbra{\R^n, \gamma_\sigma}$. We define $f$'s \emph{$\sigma$-biased Hermite coefficient of degree $k$ along $v$}, written $\wt{f}_\sigma(kv)$, to be 
	\[\wt{f}_\sigma(kv) := \Ex_{\bx\sim\calN(0,\sigma^2)^n}\sbra{f(\bx)\cdot h_{k, \sigma}\pbra{v\cdot\bx}}\]
	(as usual omitting the subscript when $\sigma=1$).
\end{definition}

\begin{notation}
	We will write $e_i \in \N^n$ to denote the $i^\text{th}$ standard basis vector for $\R^n$.
\ignore{
	\[(e_i)_j = \begin{cases}
 1 & i = j\\
 0 & i \neq j	
 \end{cases}.
\]}
\end{notation}

In this notation, for example, $\wt{f}(2e_i) = \E_{\bx\sim\calN(0,1)^n}\sbra{f(\bx)\cdot h_2(\bx_i)}$. Finally, for a measurable set $K \subseteq \R^n$, it will be convenient for us to write $\gamma(K)$ to denote $\Pr_{\bx \sim {\cal N}(0,1)^n}[x \in K]$, the \emph{(standard) Gaussian volume} of $K$.


\section{Influences for Symmetric Convex Sets}
\label{sec:influence-basics}

In this section, we first introduce our new notion of influence for symmetric convex sets over Gaussian space and establish some basic properties.  In \Cref{subsec:influence-examples} we analyze the influences of several natural symmetric convex sets, and in \Cref{subsec:russo-margulis} we give an analogue of the Margulis-Russo formula (characterizing the influences of monotone Boolean functions) which provides an alternative equivalent view of our new notion of influence for symmetric convex sets in terms of the behavior of the sets under dilations. We characterize the symmetric convex sets which have extremal max influence and total influence in \Cref{subsec:extremal}. Finally, in \Cref{subsec:other-notions-influence}, we compare our new notion of influence with some previously studied influence notions over Gaussian space.

\subsection{Definitions and Basic Properties} \label{sec:basics}

\begin{definition}[Influence for symmetric log-concave functions] \label{def:csc-influence} 
Let $f \in L^2\pbra{\R^n, \gamma}$ be a symmetric (i.e. $f(x) = f(-x)$) log-concave function. Given a unit vector $v \in \mathbb{S}^{n-1}$, we define the \emph{influence of direction $v$ on $f$} as being
\[\Inf_v[f] := -\widetilde{f}(2v) =
\Ex_{\bx \sim \calN(0,1)^n}\sbra{-f(\bx) h_{2}(v \cdot \bx)} 
= \Ex_{\bx \sim \calN(0,1)^n}\sbra{f(\bx) \cdot \pbra{\frac {1 - (v \cdot \bx)^2}{\sqrt{2}}}},\]
the negated ``degree-2 Hermite coefficient in the direction $v$.'' Furthermore, we define the \emph{total influence of $f$} as 
\[\TInf[f] := \sum_{i=1}^n \Inf_{e_i}[f].\]
\end{definition}

Note that the indicator of a symmetric convex set is a symmetric log-concave function, and this is the setting that we will be chiefly interested in. The following proposition (which first appeared in \cite{DNS20}, and a proof of which can be found in \Cref{appendix:sec-3}) shows that these new influences are indeed ``influence-like.'' An arguably simpler argument for the non-negativity of influences is presented in \Cref{subsec:russo-margulis}.

\begin{proposition}[Influences are non-negative] \label{prop:influence-nonneg}
	If $K$ is a centrally symmetric, convex set, then $\Inf_v[K] \geq 0$ for all $v\in \mathbb{S}^{n-1}$. Furthermore, equality holds if and only if $K(x) = K(y)$ whenever $x_{v^\perp} = y_{v^\perp}$ (i.e. the projection of $x$ orthogonal to $v$ coincides with that of $y$) almost surely.
\end{proposition}

We note that the total influence of a symmetric, convex set $K$ is independent of the choice of basis; indeed, we have
\begin{equation} \label{eq:total-inf-def}
	\TInf[K] = \Ex_{\bx\sim\calN(0,1)^n}\sbra{f(\bx)\pbra{\frac {n - \|\bx\|^2}{\sqrt{2}}}}	
\end{equation}
which is invariant under orthogonal transformations. Hence any orthonormal basis $\{v_1, \ldots, v_n\}$ could have been used in place of $\{e_1, \ldots, e_n\}$ in defining $\TInf[K]$. 

We note that (as is shown in the proof of \Cref{prop:influence-nonneg}), the influence of a fixed coordinate is not changed by averaging over some set of other coordinates:

\begin{fact} \label{obs:influence-averaging}
	Let $K \sse \R^n$ be a symmetric, convex set, and define the log-concave function $K_{e_i}: \R \to [0,1]$  as 
\begin{equation} \label{eq:inf-averaging-def}
	K_{e_i}(x) := \Ex_{\bx\sim\calN(0,1)^{n-1}}\sbra{K(\bx_1, \ldots, \bx_{i-1}, x,\bx_{i+1}, \ldots, \bx_n)}.	
\end{equation}
Then we have 
\begin{equation} \label{eq:inf-averaging-no-change}
	\Inf_{e_i}[K] = \Inf_{e_1}[K_{e_i}] = \TInf[K_{e_i}].
\end{equation}
\end{fact}

We conclude with the following useful relationship between the in-radius of a symmetric convex set $K$ and its max influence along any direction. 
\Cref{claim:small-inf-big-inradius} is proved in \Cref{appendix:sec-3}. 

\begin{proposition} \label{claim:small-inf-big-inradius}
Let $K \subseteq \R^n$ be a centrally symmetric convex set with $\gamma(K) \geq \Delta$, and let $\rinn=\rinn(K)$ be the in-radius of $K$. Then there is some direction $v \in \mathbb{S}^{n-1}$ such that   
\[\Inf_v[K] \geq \frac{\Delta e^{-\rinn^2}}{2^{3/2} \pi}.\]
\end{proposition}

\ignore{

}

\subsection{Influences of Specific Symmetric Convex Sets}
\label{subsec:influence-examples}

In this subsection we consider some concrete examples by analyzing the influences of a few specific  symmetric convex sets, namely ``slabs'', balls, and cubes. As we will see, these are closely analogous to well-studied monotone Boolean functions (dictator, Majority, and Tribes, respectively).

\begin{example}[Analogue of Boolean dictator: a ``slab''] \label{eg:dict}
	Given a vector $w \in \R^n$, define $\Dict_{w} := \cbra{ x \in \R^n : \abs{\abra{x, w}} \leq 1 }$. As suggested by the notation, this is the analogue of a single Boolean variable $f(x) = x_i$, i.e.~a ``dictatorship.'' For simplicity, suppose $w := \frac{1}{c}\cdot e_1$ for some $c > 0$, i.e. $\Dict_{w} = \cbra{x \in \R^n : |x_1| \leq c}$. We then have
	\[\Inf_{e_i}\sbra{\Dict_{w}} = \begin{cases}
		\Theta\pbra{c\cdot\exp\pbra{-c^2/2}} & i = 1\\
		0 & i \neq 1
	\end{cases}.\]
	Note that while in the setting of the Boolean hypercube there is only one ``dictatorship'' for each coordinate, in our setting given a particular direction we can have ``dictatorships'' of varying widths and volumes.
\end{example}

\ignore{
Recall that Majority functions are the unique maximizers of total influence among all monotone Boolean functions (see Theorem 2.33 of \cite{ODbook}). It should be clear from the following example (and, looking ahead, from \Cref{prop:ball-max-inf}) that the ball exhibits similar extremal behavior.\footnote{It is immediate from Parseval's formula and the Cauchy--Schwarz inequality that $\TInf[K] \leq \sqrt{n}$ for any symmetric convex $K\sse\R^n$, and so the ball of radius $\sqrt{n}$ has (up to constants) the largest possible total influence.} 
}
\begin{example}[Analogue of Boolean Majority: a ball] \label{eg:ball}
	Let $B_r := \cbra{ x \in \R^n : \|x\|_2 \leq r }$ denote the ball of radius $r$. Analogous to the Boolean majority function, we argue that for $B=B_{\sqrt{n}}$ we have that $\Inf_{e_i}(B)=\Theta(1/\sqrt{n})$ for all $i \in [n].$
	 
	 Recall from \Cref{eq:total-inf-def} that  
	\[\TInf\sbra{B} = \frac{1}{\sqrt{2}} \Ex_{\bx\sim\calN(0,1)^n}\sbra{B(\bx)\pbra{n - \|\bx\|^2}}.\]
By the Berry-Esseen Central Limit Theorem (see \cite{berry,esseen} or, for example, Section~11.5 of \cite{ODbook}), we have that for $t \in \R$,
\begin{equation} \label{eq:berry-essen-approx}
\left|\Prx_{\bx\sim\calN(0,1)^n}\sbra{\frac{\|\bx\|^2 - n}{\sqrt{n}} \leq t} - \Prx_{\by \sim\calN(0,1)}\sbra{\by \leq t}\right| \leq {\frac{c}{\sqrt{n}}}	\nonumber
\end{equation}
for some absolute constant $c$. 
In particular, this implies that  
\[\Prx_{\bx\sim\calN(0,1)^n}\sbra{\|\bx\|^2 \leq n - \sqrt{n}} \geq 
\Prx_{\by \sim {\cal N}(0,1)}[\by \leq -1] -
\frac{c}{\sqrt{n}} \geq 0.15.\]
Since $\Pr_{\bx \sim {\cal N}(0,1)^n}[B(\bx)=1] = \frac12 \pm o_n(1),$ and $B(x)(n-\|x\|^2)$ is never negative, it follows that
\[\Ex_{\bx\sim\calN(0,1)^n}\sbra{B(\bx)\pbra{n - \|\bx\|^2}} \geq \Theta\pbra{\sqrt{n}}\]
from which symmetry implies that $\Inf_{e_i}\sbra{B} \geq \Theta\pbra{\frac{1}{\sqrt{n}}}$ for all $i \in [n].$ The upper bound $\Inf_{e_i}\sbra{B} \leq \Theta\pbra{\frac{1}{\sqrt{n}}}$ follows from Parseval's identity.
\end{example}

Our last example is analogous to the ``Tribes CNF'' function introduced by Ben-Or and Linial \cite{BenOrLinial:85short} (alternatively, see Definition 2.7 of \cite{ODbook}):

\begin{example}[Analogue of Boolean $\Tribes$: a cube] \label{eg:solid-cube}
	Let $C_r := \cbra{ x \in \R^n : |x_i| \leq r \text{ for all } i \in [n]}$ denote the axis-aligned cube of side-length $2r$ and $\gamma(C_{r}) = \frac{1}{2}$, i.e. let $r > 0$ be the unique value such that 
	\begin{equation} \label{eq:bbb}\Prx_{\bg\sim\calN(0,1)}\sbra{|\bg| \leq r} = \pbra{\frac{1}{2}}^{1/n} = 1 - \frac{\Theta(1)}{n}.\end{equation}
	By standard tail bounds on the Gaussian distribution, we have that $r = \Theta(\sqrt{\log n})$. 
	Because of the symmetry of $C_r$, we have 
	$\Inf_{e_i}[C_{r}] = \Inf_{e_j}[C_{r}]$ for all $i, j \in [n]$. Note, however, that we can write 
	\[C_{r}(x) = \prod_{i=1}^n \Dict_{1/r}(x_i)\]
	where $\Dict_{1/r} : \R \to \zo$ is as defined in \Cref{eg:dict}. By considering the Hermite representation of $C_r(x),$ it is easy to see that 
	\[\Inf_{e_i}[C_r] = \Ex_{\bg\sim\calN(0,1)}\sbra{\Dict_{1/r}(\bg)}^{n-1}\TInf\sbra{\Dict_{1/r}}.\] 
	By our choice of $r$ above, we have $\E\sbra{\Dict_{1/r}} = \sqrt[n]{1/2}$ and so 
	\[ \Ex_{\bg\sim\calN(0,1)}\sbra{\Dict_{1/r}(\bg)}^{n-1} = \Theta(1). \]
	From \Cref{eg:dict}, we know $\TInf\sbra{\Dict_{1/r}} = \Theta\pbra{r e^{-r^2/2}}$, and so we have
	\begin{equation} \label{eq:aaa}\Inf_{e_i}[C_r] = \Theta\pbra{re^{-r^2/2}}.
	\end{equation}
	We now recall the following tail bound on the normal distribution (see Theorem~1.2.6 of \cite{durrett_2019} or Equation~2.58 of \cite{TAILBOUND}):
\begin{equation} \label{eq:normal-tail}
\varphi(r)
\left({\frac 1 r} - {\frac 1 {r^3}} \right) \leq \Prx_{\bg \sim N(0,1)}[\bg \geq r] \leq
\varphi(r)
\left({\frac 1 r} - {\frac 1 {r^3}} + {\frac 3 {r^5}}\right),
\end{equation}
where $\varphi(r) = {\frac 1 {\sqrt{2 \pi}}} e^{-r^2/2}$ is the density function of $N(0,1)$.
Combining \Cref{eq:bbb}, \Cref{eq:aaa} and \Cref{eq:normal-tail} we get that
$
\Inf_{e_i}[C_r] = \Theta(r^2) \cdot \Prx_{\bg \sim N(0,1)}[\bg \geq r] = \Theta(\log(n)) \cdot \Theta(1/n),
$
which corresponds to the influence of each individual variable on the Boolean ``tribes'' function.

\ignore{	
}	
\end{example}

\subsection{Margulis-Russo for Convex Influences:  An Alternative Characterization of Influences via Dilations}
\label{subsec:russo-margulis}

In this subsection we give an alternative view of the notion of influence defined above, in terms of the behavior of the Gaussian measure of the set as the variance of the underlying Gaussian is changed.\footnote{Since $\gamma_\sigma(K) = \gamma(K/\sigma)$, decreasing (respectively increasing) the variance of the underlying Gaussian measure is equivalent to dilating (respectively shrinking) the set.} This is closely analogous to the Margulis-Russo formula  for monotone Boolean functions on $\bn$ (see \cite{Russo:81, Margulis:74} or Equation~(8.9) in \cite{ODbook}), which relates the derivative (with respect to $p$) of the $p$-biased measure of a monotone function $f$ to the $p$-biased total influence of $f$.  

We start by defining $\sigma$-biased convex influences, which are analogous to $p$-biased influences from Boolean function analysis (see Section~8.4 of \cite{ODbook}). 

\begin{definition}[$\sigma$-biased influence] \label{def:sigma-biased-influence}
	Given a centrally symmetric convex set $K \sse \R^n$, we define the \emph{$\sigma$-biased influence of direction $v$ on $K$} as being
\[\Inf^{(\sigma)}_v[K] := -\widetilde{f}_\sigma(2v) =
\Ex_{\bx \sim \calN(0,1)^n}\sbra{-f(\bx) h_{2, \sigma}(v \cdot \bx)},\]
the negated degree-2 $\sigma$-biased Hermite coefficient in the direction $v$. We further define the \emph{$\sigma$-biased total influence of $K$} as 
\[\TInf^{(\sigma)}[K] := \sum_{i=1}^n \Inf_{e_i}^{(\sigma)}[K].\]
\end{definition}

The proof of the following proposition, which asserts that the rate of the change of the Gaussian measure of a symmetric convex set $K$ with respect to $\sigma^2$ is (up to scaling) equal to the $\sigma$-biased total influence of $K$, is deferred to \Cref{appendix:sec-3}. We note that this relation was essentially known to experts (see e.g.~\cite{s-inequality}), though we are not aware of a specific place where it appears explicitly in the literature.

\begin{proposition}[Margulis-Russo for symmetric convex sets]
\label{prop:russo-margulis-convex}
	Let $K \sse \R^n$ be a centrally symmetric convex set. Then for any $\sigma > 0$ we have
	\[\frac{d}{d\sigma^2} \E_{\bx\sim\calN(0,\sigma^2)^n}\sbra{K(\bx)} = \frac{-\TInf^{(\sigma)}[K]}{\sigma^2\sqrt{2}}
	= \frac{-1}{\sigma^2\sqrt{2}}\sum_{i=1}^n \Inf^{(\sigma)}_{e_i}[K].\]
\end{proposition}

Note that decreasing (respectively increasing) the variance of the background Gaussian measure is equivalent to dilating (respectively shrinking) the symmetric convex set while keeping the background measure fixed; this lets us write\ignore{
\rnote{Does this need a ${\frac 1 {\sqrt{2}}}$ in front?  It seems to me that
\begin{align*}
\TInf[K] &= \sqrt{2} 
\frac{d}{d\sigma^2} \Ex_{\bx\sim\calN(0,\sigma^2)^n}\sbra{K(\bx)} \bigg|_{\sigma^2 = 1}\\
&=\sqrt{2} 
\lim_{h \to 0}
{\frac {\Ex_{\bx\sim\calN(0,1+h)^n}\sbra{K(\bx)} - \Ex_{\bx\sim\calN(0,1)^n}\sbra{K(\bx)}}
h}
&=\sqrt{2} 
\lim_{h \to 0}
{\frac {\Ex_{\bx\sim\calN(0,1)^n}\sbra{K(\bx/\sqrt{1+h})} - \Ex_{\bx\sim\calN(0,1)^n}\sbra{K(\bx)}}
h}\\
&=\sqrt{2} 
\lim_{h \to 0}
{\frac {\Ex_{\bx\sim\calN(0,1)^n}\sbra{K(\bx(1-h/2))} - \Ex_{\bx\sim\calN(0,1)^n}\sbra{K(\bx)}}
h}\\
&=\sqrt{2} 
\lim_{h \to 0}
{\frac {\Ex_{\bx\sim\calN(0,1)^n}\sbra{K(\bx(1-h))} - \Ex_{\bx\sim\calN(0,1)^n}\sbra{K(\bx)}}
{2h}}\\
&= {\frac 1 {\sqrt{2}}}
\lim_{\delta \to 0}
{\frac {\Ex_{\bx\sim\calN(0,1)^n}\sbra{K(\bx(1-\delta))} - \Ex_{\bx\sim\calN(0,1)^n}\sbra{K(\bx)}}
{\delta}}
\end{align*}
If this is correct let's just put the corrected formula without this derivation.
}}
\begin{equation}
\label{eq:russo-margulis-dilations}
	\TInf[K] ={\frac 1 {\sqrt{2}}} \lim_{\delta\to0}\frac{\gamma_n(K) - \gamma_n\pbra{(1-\delta)K}}{\delta}
\end{equation}
for a symmetric convex $K\sse\R^n$. We also note that \Cref{prop:russo-margulis-convex} easily extends to the following coordinate-by-coordinate version (which also admits a similar description in terms of dilations): 

\begin{proposition}[Coordinate-wise Margulis-Russo]
\label{prop:coordinate-russo-margulis}
	Let $K \sse \R^n$ be a centrally symmetric convex set. Then for any $\sigma > 0$, we have 
	\[
	{\frac {d}{d\sigma_i^2}}
 \Ex_{\substack{\bx_i \sim \calN(0,\sigma_i^2)\\j\neq i \ :\  \bx_j\sim\calN(0,\sigma^2)}} \sbra{K(\bx)} \bigg|_{\sigma_i^2 = \sigma^2}	
= \frac{-1}{\sigma^2\sqrt{2}}\Inf^{(\sigma)}_{e_i}[K].
 \]
\end{proposition} 

In particular, we have
\[\Inf_{e_i}[K] = -\sqrt{2}\frac{d}{d\sigma^2} \Ex_{\substack{\bx_i \sim \calN(0,\sigma^2)\\j\neq i \ :\  \bx_j\sim\calN(0,1)}} \sbra{K(\bx)} \bigg|_{\sigma^2 = 1}.\] 
Note that decreasing the variance of the underlying Gaussian measure along a coordinate direction cannot cause the volume of the set to decrease.\ignore{\rnote{Not sure we need the next footnote - isn't this just a consequence of the fact that decreasing the variance of the Gaussian in a coordinate direction is equivalent to stretching/dilating the body in that direction while leaving the measure unchanged, and clearly this can't decrease volume? If so can we delete the next footnote?}\footnote{This is easy to see from the fact that the indicator of a symmetric convex set is a unimodal log-concave function, and that marginals of log-concave functions are log-concave (see \Cref{fact:marginal-log-concave}).}} It follows then that $\Inf_{e_i}[K] \geq 0$ for all $e_i$. 


\subsection{Extremal Symmetric Convex Sets}
\label{subsec:extremal}


The unique maximizer of $\Inf_1[f]$ across all monotone Boolean functions $f: \bn \to \bits$ is the dictator function $f(x)=x_1$.  The next proposition gives an analogous statement for the ``dictatorship'' function $\Dict_{w}$ from \Cref{eg:dict}, for every possible Gaussian volume:

\ignore{The next proposition is analogous to the easy-to-see statement in the setting of monotone Boolean functions on the hypercube that for any monotone $f \isafunc$, we have $\Inf_i[f] \leq \Inf_i[x_i] = 1$. }

\begin{proposition}
\label{prop:dict-maximizes-inf-along-coordinate}
	Let $K \sse \R^n$ be a symmetric convex set and let $v \in \mathbb{S}^{n-1}$. Let $c \geq 0$ be chosen so that the ${\cal N}(0,1)^n$ Gaussian volume of $\Dict_{cv}$ equals that of $K$, i.e.~$\gamma(\Dict_{cv}) = \gamma(K)$. Then $\Inf_{v}[K] \leq \Inf_{v}[\Dict_{cv}]$. 	
\end{proposition}

\begin{proof}
Without loss of generality (for ease of notation) we take $v=e_1$. 
Let $g_K: \R \to [0,1]$ be the function obtained by marginalizing out variables $x_2,\dots,x_n$, so 
\[
g_K(x_1) = \Ex_{(\bx_2,\dots,\bx_n) \sim {\cal N}(0,1)^{n-1}}\sbra{K(x_1,\bx_2,\dots,\bx_n)}.
\]
As noted following\ignore{\rnote{Do we want to package this into an observation with a number for subsequent reference (like here)?}}  \Cref{prop:influence-nonneg}, we have that
$\Inf_{e_1}[K]=\Inf_{e_1}[g_K]$. We observe that by definition we have
\[
\Inf_{e_1}[g_K] = {\frac 1 {\sqrt{2}}} \cdot \Ex_{\bx_1 \sim {\cal N}(0,1)}\sbra{(1 - \bx_1^2)g_K(\bx_1)}.
\]
Since $1-t^2$ is a decreasing function of $t$ for all $t \geq 0$, it is easy to see that the symmetric $[0,1]$-valued function $g_K$ that maximizes $\Ex_{\bx_1 \sim {\cal N}(0,1)}\sbra{(1 - \bx_1^2)g_K(\bx_1)}$ subject to having  $\E_{\bx_1 \sim {\cal N}(0,1)}[g_K(\bx_1)]=\gamma(\Dict_{ce_1})$ is the function $g$ for which $g(t)=1$ for $|t| \leq c$ and $g(t)=0$ for $|t| > c$. This corresponds precisely to having $K = \Dict_{ce_1}$; so in fact taking $K=\Dict_{ce_1}$ maximizes $\Inf_{e_1}[K]$ over all measurable subsets of $\R^n$ of Gaussian volume $\gamma\pbra{\Dict_{ce_1}}$ (not just over all symmetric convex sets of that volume). 
\end{proof}

We note that a slight extension of this argument can be used to give a robust version of \Cref{prop:dict-maximizes-inf-along-coordinate}, showing that for any $c>0$, any symmetric convex set $K$ (in fact any measurable set $K$) of Gaussian volume $\gamma(\Dict_{cv})$ that has $\Inf_{v}[K]$ close to $\Inf_v[\Dict_{cv}]$ must in fact be close to $\Dict_{cv}$. This is analogous to the easy fact that any monotone Boolean function with $\Inf_1[f]$ close to 1 must be close to the function $f(x)=x_1$.

Next we give a similar result but for total convex influence rather than influence in a single direction, analogous to the well known fact that the Majority function maximizes total influence across all $n$-variable monotone Boolean functions $f: \bn \to \bits$:
\ignore{

}
\begin{proposition}
\label{prop:ball-max-inf}
	Let $K \sse \R^n$ be a symmetric convex set, and let $r \geq 0$ be chosen so that the ${\cal N}(0,1)^n$ Gaussian volume of $B_r$ equals that of $K$, i.e.~$\gamma(B_r) = \gamma(K)$. Then $\TInf[K] \leq \TInf[B_r]$.
\end{proposition}

\begin{proof}
The argument is similar to that of \Cref{prop:dict-maximizes-inf-along-coordinate}.
	We have 
	\[\TInf[K] = \Ex_{\bx\sim\calN(0,1)^n}\sbra{K(\bx)\pbra{\frac {n - \|\bx\|_2^2}{\sqrt{2}}}}
	= {\frac 1 {\sqrt{2}}}\cdot \Ex_{\br \sim \chi(n)}\sbra{(n-\br^2)\alpha_K(\br)}
	\] 
(recall \Cref{eq:sdf}), where $\chi(n)$ is the $\chi$-distribution with $n$ degrees of freedom. We observe that taking $K=B_r$ results in $\alpha_K(t) = 1$ for $t \leq r$ and $\alpha_K(t)=0$ for $t>r$, that the range of $\alpha_K(\cdot)$ is contained in $[0,1]$ for any $K$, and that $n-t^2$ is a decreasing function of $t$ for all $t \geq 0$. Combining these observations, it is easily seen that taking $K=B_r$ in fact maximizes the expression on the RHS over all measurable subsets of $\R^n$ of volume $\gamma(B_r)$ (not just over all symmetric convex sets of that volume).
\end{proof}

As before, the argument above can be used to establish a robust version of \Cref{prop:ball-max-inf}, showing that any symmetric convex set $K$ (in fact any measurable set $K$) of Gaussian volume $\gamma(B_r)$ that has $\TInf[K]$ close to $\TInf[B_r]$ must in fact be close to $B_r$.


\subsection{Other Notions of Influence} 
\label{subsec:other-notions-influence}

Here, we compare the notion of influence for symmetric convex sets proposed in \Cref{def:csc-influence} with two previous notions of influence, namely i) the \emph{geometric influence} introduced in \cite{keller2012geometric}; and ii) the \emph{expected variance along a fiber} which coincides with the usual notion of influence for Boolean functions on the hypercube. 

\subsubsection{Geometric Influences}
\label{subsubsec:geometric-inf}

In \cite{keller2012geometric}, Keller, Mossel, and~Sen introduced the notion of \emph{geometric influence} for functions over Gaussian space, and proved analogues of seminal results from the analysis of Boolean functions---including the KKL theorem, the Margulis--Russo lemma, and an analogue of Talagrand's correlation inequality---for this notion of influence. Informally, the geometric influence captures the expected \emph{lower Minkowski content} along each one-dimensional fiber of a set. 

\begin{definition}[Geometric influences]	 \label{def:kms-geometric-influences}
	Given a Borel measurable set $K\sse\R$, its \emph{lower Minkowski content} (with respect to the standard Gaussian measure), denoted $\gamma^+$, is defined as
	\[\gamma^+(K) := \lim\inf_{r \to 0^+} \frac{\gamma\pbra{K + [-r, r]} - \gamma(K)}{r}.\]
(Note that for $K = [a, b] \subset \R$, we have 
$\gamma^{+}(K) = \varphi(a) + \varphi(b).$)
	For any Borel-measurable $K \sse\R^n$, for each $i \in [n]$ and $x \in \R^n$, define the fiber
	\[K^x_i := \{y\in\R : (x_1, \ldots, x_{i-1}, y, x_{i+1}, \ldots, x_n) \in  K\}.\] 
	The \emph{geometric influence of coordinate $i$ on $K$} is 
	\[\GInf_i[K] := \Ex_{\bx\sim\calN(0,1)^n}\sbra{\gamma^+(K^{\bx}_i)}.\]
\end{definition}

For convex sets the \emph{total geometric influence} admits a geometric interpretation as the change in the boundary of the set under uniform enlargement: 

\begin{proposition}[Remark 2.2 of \cite{keller2012geometric}] 
\label{prop:geom-meaning-geom-inf}
Let $K\sse\R^n$ be a convex set. Then we have 
\[\lim_{r \to 0^+} \frac{\gamma_n(K + [-r, r]^n) - \gamma_n(K)}{r} = \sum_{i=1}^n \Inf_i^{\calG}[K]\]
	where the right-hand side above is the \emph{total geometric influence of $K$}.
\end{proposition}

Note that unlike $\TInf[K]$ (see \Cref{def:csc-influence}), the total geometric influence is not invariant under rotations as the boundary of a convex set under enlargement is not rotationally invariant.  It is possible for our convex influence notion  to be much smaller than the geometric influence. For example, a routine computation shows that 
the $\sqrt{n}$-radius Euclidean ball $B := B_{\sqrt{n}}$ has $\GInf_{i}[B] = \Omega(1)$ for each $i \in [n]$, whereas as seen from \Cref{eg:ball}, for convex influence we have $\Inf_{e_i}[B] = O\pbra{\frac{1}{\sqrt{n}}}$. 

\ignore{

for  ( the {influence} of a coordinate on a symmetric convex set to be much smaller than the {geometric influence} of that coordinate on the set.

The following example shows that it is possible for the {influence} of a coordinate on a symmetric convex set to be much smaller than the {geometric influence} of that coordinate on the set. 

\begin{example}
\label{eg:geom-inf-vs-dns-inf}

Consider $B := B_{\sqrt{n}} \sse\R^n$ as defined in \Cref{eg:ball}. 
For all $x \in \R^{n}$, we have $B^x_i = [-a, a]$ (where $B^x_i$ is as defined in \Cref{def:kms-geometric-influences}) for some $a$. Let $\bx \sim \calN(0,1)^n$; recall that $\by^2 := \sum_{i=2}^n \bx_i^2$ is distributed as $\by^2\sim\chi^2(n-1)$, and so we have from standard tail bounds on $\chi^2(n-1)$ that $\Pr\sbra{\bx_1^2 \leq 1} = \Omega(1)$,  and so $\GInf_1[B] = \Omega(1)$. On the other hand, as seen from \Cref{eg:ball}, $\Inf_{e_1}[B] = O\pbra{\frac{1}{\sqrt{n}}}$. 
\end{example}

\red{TODO: Comparison?}
}


\subsubsection{Variance Along a Fiber}
\label{subsubsec:varinf}

In the setting of  Boolean functions over the hypercube, the usual notion of influence of a coordinate on a function $f: \bn \to \bits$ coincides with the expected variance of the function along a random fiber in the direction of that coordinate. This is also a standard notion of influence for product probability measures more generally, see e.g. Proposition~8.24 of \cite{ODbook}.  More formally, we have the following definition. 

\begin{definition}[Expected variance along a fiber] \label{def:varinf}
	Given a function $f \in L^2\pbra{\R^n, \gamma}$, we define
	\[\VarInf_i[f] := \Ex_{\bx\backslash\{\bx_i\} \sim \calN(0,1)^{n-1}}\sbra{\Var_{\bx_i}[f(\bx)]}\]
	to be the \emph{expected variance of $f$ along the $i^\text{th}$ fiber}. 
\end{definition}


We can express the expected variance along the $i^\text{th}$ fiber in terms of the Hermite expansion as $\VarInf_i[f] = \sum_{\alpha_i > 0} \wt{f}(\alpha)^2$ (see Proposition 8.23 of \cite{ODbook}).  In our setting it is possible for the convex influence of a symmetric convex set to be much smaller than the expected variance along a fiber. This is witnessed by the symmetric convex set $\Dict_v \sse\R^n $ given by 
\[\Dict_v := \{x \in \R^n : |x\cdot v| \leq 1\} \qquad\text{where}\qquad v = \frac{1}{\sqrt{n}}(1, \ldots, 1).\] 
A routine computation shows that $\VarInf_i\sbra{\Dict_v} = \Theta\pbra{\frac{1}{\sqrt{n}}}$ for each $i \in [n]$. On the other hand,  since $\TInf[\Dict_v]$ is rotationally invariant, it follows from \Cref{eg:dict} that $\TInf\sbra{\Dict_v} = \TInf\sbra{\Dict_{e_1}} = \Theta(1)$, and consequently by symmetry, it follows that $\Inf_{e_i}[K] = \Theta\pbra{\frac{1}{n}}$.

\section{Lower Bounds on Total Convex Influence}
\label{sec:poincare-kkl}

Two fundamental results on the influence of variables for Boolean functions $f: \bn \to \bits$ are the Poincar\'e inequality and the celebrated ``KKL Theorem'' of Kahn, Kalai, and Linial \cite{KKL:88}, both of which give lower bounds on total influence. The former states that the total influence of any $f: \bn \to \bits$ is at least its variance, and has a very elementary proof (indeed it can be proved in a single line by comparing the Fourier expressions for the two quantities).   The KKL  theorem gives a more refined bound, showing (roughly speaking) that if all influences are small then the total influence must be somewhat large. Several proofs of the KKL Theorem are now known, using a range of different techniques such as  the famous hypercontractive inequality \cite{Bon70,Bec75} (the original approach), methods of stochastic calculus \cite{EldanGross:20}, and the Log-Sobolev inequality \cite{KKKMS:21}.

In this section we prove convex influence analogues of the Poincar\'{e} inequality and the KKL theorem.
We use the ``$S$-inequality'' of Lata\l a and Oleskiewicz \cite{s-inequality} to give a relatively quick proof our Poincar\'{e} analogue, and prove our analogue of the KKL theorem using the Gaussian Isoperimetric Theorem \cite{Borell:75}.
 
\subsection{A Poincar\'e Inequality for Symmetric Convex Sets}
\label{subsec:poincare}

Recall from \Cref{subsec:other-notions-influence} that our convex influence notion can be much smaller than the geometric influence defined in \cite{keller2012geometric} or the ordinary ``variance along a fiber'' influence notion of \Cref{subsubsec:varinf}. Given this, it is of interest to give \emph{lower bounds} on the total convex influence.  Our first result along these lines is an analogue of the standard Poincar\'{e} inequality for Boolean functions (or more generally for functions over product domains):

\begin{proposition}[Poincar\'e for symmetric convex sets]
\label{prop:poincare-csc-sets}
	Let $K\sse\R^n$ be a symmetric convex set. Then 
$\TInf[K] \geq \Omega\pbra{\Var[K]}.$
\end{proposition}

The main tool we use for our proof of \Cref{prop:poincare-csc-sets} is the following celebrated result of Lata\l a and Oleskiewicz concerning the rate of growth of symmetric convex sets under dilations: 

\begin{proposition}[$S$-inequality, Theorem 1 of \cite{s-inequality}]
\label{prop:s-inequality}
	Let $K\sse\R^n$ be a symmetric convex set, and let $\Dict_{w}\sse\R^n$ be a symmetric strip (i.e. $\Dict_w = \cbra{x \in \R^n : |x\cdot w| \leq 1}$ for some fixed $w \in \R^n$) such that $\gamma_n(A) = \gamma_n\pbra{\Dict_w}$. Then 
	\[\gamma_n(tK) \geq \gamma_n\pbra{\Dict_{w/t}} \qquad \text{for } t\geq 1,\]
	and 
	\[\gamma_n(tK) \leq \gamma_n\pbra{\Dict_{w/t}} \qquad \text{for } 0\leq t\leq 1.\] 
\end{proposition}

Intuitively, the above result says that among all convex symmetric sets of a given Gaussian volume, the Gaussian volume of dictatorships (see \Cref{eg:dict}) grows the slowest under enlargement by dilations. 

The proof of \Cref{prop:poincare-csc-sets} combines \Cref{prop:s-inequality} with our Margulis-Russo analogue (the characterization of influence in terms of dilations given in \Cref{subsec:russo-margulis}):

\medskip

\begin{proofof}{\Cref{prop:poincare-csc-sets}}
	Write $\gamma_n(K) = \alpha$, and let $\Dict_{e_1/a} = \{x\in\R^n : |x_1| \leq a\}$ be such that $\gamma_1\pbra{\Dict_{e_1/a}} = \alpha$, i.e. $\gamma_n\pbra{[-a, a]} = \alpha$. Recall from \Cref{subsec:russo-margulis} that 
	\begin{equation}
	\label{eq:basic-inf-def}	
	\TInf[K] = {\frac 1 {\sqrt{2}}} \lim_{\delta\to0} \frac{\gamma_n(K) - \gamma_n\pbra{(1-\delta)K}}{\delta}.
	\end{equation}
	By the $S$-inequality (\Cref{prop:s-inequality} above), for any fixed $0 < \delta \leq 1$, we have 
	\[\gamma_n\pbra{(1-\delta)K} \leq \gamma_n\pbra{(1-\delta)\Dict_{e_1/a}} = \gamma_1\pbra{\sbra{-(1-\delta)a, (1-\delta)a}},\]
	which implies that
	\begin{align}
		\gamma_n(K) - \gamma_n\pbra{(1-\delta)K} &\geq \gamma_1\pbra{[-a, a]} - \gamma_1\pbra{\sbra{-(1-\delta)a, (1-\delta)a}} \nonumber \\
		&= \frac{1}{\sqrt{2\pi}}\pbra{\int_{-a}^a e^{-x^2/2}\,dx - \int_{-(1-\delta)a}^{(1-\delta)a} e^{-x^2/2}\,dx} \nonumber \\
		&= \sqrt{\frac{2}{\pi}}\int_{(1-\delta)a}^a e^{-x^2/2}\, dx. \label{eq:final-step-poincare}
	\end{align}
When $\alpha \leq 1/2$, we have $\alpha \leq a \leq 1$, and clearly
	\begin{equation}
	\label{eq:poincare-case-1}	
	\sqrt{\frac{2}{\pi}}\int_{(1-\delta)a}^a e^{-x^2/2}\, dx \geq \Omega\pbra{\delta a} \geq \Omega\pbra{\delta \alpha}.
	\end{equation}
	Combining \Cref{eq:basic-inf-def,eq:final-step-poincare,eq:poincare-case-1} implies the desired result in this case. On the other hand, when $\alpha \geq 1/2$, we have 
	\[\sqrt{\frac{2}{\pi}}\int_{(1-\delta)a}^a e^{-x^2/2}\, dx \geq \Omega\pbra{\delta a e^{-a^2/2}}.\]
	Standard tail bounds on the Gaussian distribution give $ae^{-a^2/2} \geq \Omega\pbra{1 - \Phi(a) }$ when $a \geq 1$ (see, for example, Theorem 1.2.6 of \cite{durrett_2019}). It follows that if $\gamma_1\pbra{[-a, a]} = \frac{1}{\sqrt{2\pi}}\int_{-a}^a e^{-x^2}\,dx = \alpha$, then $ae^{-a^2/2} \geq \Omega(1-\alpha)$. In particular, we have
	\begin{equation}
	\label{eq:poincare-case-2}	
	\sqrt{\frac{2}{\pi}}\int_{(1-\delta)a}^a e^{-x^2/2}\, dx \geq \Omega\pbra{\delta(1-\alpha)}
	\end{equation}
	and combining \Cref{eq:basic-inf-def,eq:final-step-poincare,eq:poincare-case-2} implies the desired result. 
\end{proofof}

\medskip

\subsection{A KKL Analogue for Symmetric Convex Sets}
\label{subsec:kkl}

For the symmetric convex set $\Dict_{e_1}$, both the total convex influence and the variance  are $\Theta(1)$, so \Cref{prop:poincare-csc-sets} is best possible (up to a constant factor) for arbitrary symmetric convex sets.  But of course $\Dict_{e_1}$ has very large (constant) influence in a single direction, analogous to a Boolean function with an individual coordinate of constant influence. The famous KKL theorem for Boolean functions over $\bn$ states that if no  coordinate influence is allowed to be large (each is at most $\delta$),  then the total influence must be large (at least $\Omega(\Var[f] \cdot \log(1/\delta))$).  We now prove an analogous result for convex influences, though we only achieve a quadratically weaker bound in terms of the max influence:

\begin{theorem}[KKL for symmetric convex sets]
\label{thm:our-kkl}
	Let $K\sse\R^n$ be a symmetric convex set with $\Inf_v[K] \leq \delta \leq \Var[K]/10$ for all $v \in \mathbb{S}^{n-1}$. Then
	\begin{equation}
		\label{eq:kkl-convex-sets}
\TInf[K] \geq \Omega\pbra{\Var[K]\sqrt{\log\pbra{\frac{\Var[K]}{\delta}}}}.
	\end{equation}
\end{theorem}

Our proof of \Cref{thm:our-kkl} is inspired by the approach of \cite{lat-ole-kkl}.  The main technical ingredient we use is the \emph{Gaussian isoperimetric inequality:}

\begin{proposition}[Gaussian isoperimetric inequality, \cite{Borell:75}]
\label{prop:gaussian-isoperimetric-inequality}
	Given any Borel set $A \sse\R^n$, we have 
	\[\Phi^{-1}\pbra{\gamma_n\pbra{A_t}} \geq \Phi^{-1}\pbra{\gamma_n\pbra{A}} + t\] 
	where $A_t := A + B_t$ is the $t$-enlargement of $A$.
\end{proposition}

We remark that it is easy to obtain \Cref{prop:gaussian-isoperimetric-inequality} from the Ehrhard-Borell inequality \cite{Ehrhard:83,Borell:03,Borell:08}, which we recall as \Cref{prop:ehrhard-borell} in \Cref{appendix:sec-3}. We will also require the following easy estimate on the \emph{Gaussian isoperimetric function} $\varphi\circ\Phi^{-1}(\cdot)$.

\begin{proposition} \label{prop:gaussian-isop-func-estimate}
	Let $\Phi: \R \to [0, 1]$ denote the cumulative distribution function of the standard one-dimensional Gaussian distribution, and let $\varphi := \Phi'$ denote its density. Then for all $\alpha \in (0, 1)$, we have
	\[\varphi\circ\Phi^{-1}(\alpha) \geq \sqrt{\frac{2}{\pi}}\min(\alpha, 1-\alpha).\]
\end{proposition}

\begin{proof}
	By symmetry, it suffices to show that $\varphi\circ\Phi^{-1}(\alpha) \geq \sqrt{\frac{2}{\pi}}\alpha$ for $\alpha\in\sbra{0,\frac{1}{2}}$. This is immediate from the fact that 
	\[\varphi\circ\Phi^{-1}(0) = 0 \qquad\text{and}\qquad \varphi\circ\Phi^{-1}\pbra{\frac{1}{2}} = \frac{1}{\sqrt{2\pi}},\] 
	and the concavity of $\varphi\circ\Phi^{-1}$ (see, for example, Exercise~5.43 of \cite{ODbook}).
\end{proof}

\medskip

\begin{proofof}{\Cref{thm:our-kkl}}
	Let $\rinn$ denote the in-radius of $K$. We will show that 
	\begin{equation}
		\label{eq:kkl-first-goal}
		\TInf[K] \geq {\frac 1 {\sqrt{\pi}}}\Var\sbra{K}\cdot\rinn
	\end{equation}
	and that
	\begin{equation} \label{eq:kkl-second-goal}
	\rinn \geq \Omega(\sqrt{\ln(\Var[K]/\delta)})
	\end{equation}
	from which the desired result follows. 
	
	For \Cref{eq:kkl-second-goal}, by \Cref{claim:small-inf-big-inradius} we have that for some direction $v \in \mathbb{S}^{n-1}$,
\[
\Inf_{\hat{v}}[K] \geq {\frac {\gamma(K) e^{-\rinn^2}}{2^{3/2} \pi}}
\geq {\frac {\Var[K] e^{-\rinn^2}}{2^{3/2} \pi}}.
\]
	Combining this with $\Inf_{\hat{v}}[K] \leq \delta$\ignore{ (and observing that we may assume $\delta$ is at most some sufficiently small absolute constant, since otherwise \Cref{thm:our-kkl} is implied by \Cref{prop:poincare-csc-sets})} and recalling that $\delta \leq \Var[K]/10$, we get \Cref{eq:kkl-second-goal}.

	\ignore{We begin with \Cref{eq:kkl-second-goal}: by definition of the in-radius, for any $\zeta>0$, there is a point $z_\ast$ such that $z_\ast \not \in K$ and $\Vert z_\ast \Vert_2 = \rinn + \zeta$. 
By the separating hyperplane theorem, there must exist
some unit vector $\hat{v} \in \mathbb{R}^n$ such that 
\begin{equation}~\label{eq:inclusion-0}
K \subseteq  K_{\ast} \coloneqq \{x \in \mathbb{R}^n: |\hat{v} \cdot x| \le \rinn+ \zeta\}, \nonumber
\end{equation}
and hence by \Cref{claim:small-inf-large-slab} we get that
\[
\Inf_{\hat{v}}[K] \geq {\frac {\gamma(K) e^{-(\rinn + \zeta)^2}}{2^{3/2} \pi}}
\geq {\frac {\Var[K] e^{-(\rinn + \zeta)^2}}{2^{3/2} \pi}}.
\]
Combining this with $\Inf_{\hat{v}}[K] \leq \delta$ (and observing that we may assume $\delta$ is at most some sufficiently small absolute constant, since otherwise \Cref{thm:our-kkl} is implied by \Cref{prop:poincare-csc-sets}), letting $\zeta \to 0^+$ we get \Cref{eq:kkl-second-goal}.}

We turn now to establishing \Cref{eq:kkl-first-goal}.
	Recall from \Cref{eq:russo-margulis-dilations} of \Cref{subsec:russo-margulis} (our Margulis-Russo formula) that  
	\begin{equation}
		\label{eq:kkl-inf-def}
		\TInf[K] ={\frac 1 {\sqrt{2}}} \lim_{\delta\to0} \frac{\gamma(K) - \gamma\pbra{(1-\delta)K}}{\delta}.
	\end{equation}
	We proceed to upper-bound $\gamma\pbra{(1-\delta)K}$ in terms of $\gamma(K)$. Since $\rinn$ is the in-radius of $K$, for all $0 < \delta \leq 1$, we have that
	\begin{equation} \label{eq:kkl-containment}
		(1-\delta)K + \delta \rinn B_{1} = (1-\delta)K + B_{\delta \rinn} \sse K.
	\end{equation}
 Let $K^c := \R\backslash K$, and let $(K^c)_{\delta \rinn} := K^c + B_{\delta \rinn}$ be the $\delta \rinn$-enlargement of $K^c$. It follows from \Cref{eq:kkl-containment} that $(1-\delta)K\cap (K^c)_{\delta \rinn} = \emptyset$, which in turn implies that
	\begin{equation} \label{eq:kkl-first-ub}
	\gamma((1-\delta)K) + \gamma((K^c)_{\delta\rinn}) \leq 1, 
	\qquad\text{and so}\qquad
 	\gamma((1-\delta)K) \leq 1 - \gamma((K^c)_{\delta\rinn}).
	\end{equation}
	However, from the Gaussian isoperimetric inequality (\Cref{prop:gaussian-isoperimetric-inequality}), we know that 
	\begin{equation} \label{eq:kkl-gip}
		\gamma\pbra{(K^c)_{\delta\rinn}} \geq \Phi\pbra{\Phi^{-1}\pbra{\gamma\pbra{K^c}} + \delta\rinn}. 
	\end{equation} 
Let $\alpha = \gamma\pbra{K^c}$, so $\gamma(K) = 1-\alpha$. Putting \Cref{eq:kkl-inf-def,eq:kkl-first-ub,eq:kkl-gip} together, we get 
	\begin{align*}
	\TInf[K] &\geq {\frac 1 {\sqrt{2}}} \lim_{\delta\to0} \frac{\Phi\pbra{\Phi^{-1}\pbra{\alpha} + \delta\rinn} - \alpha}{\delta}\\
	&= {\frac 1 {\sqrt{2}}} \rinn\pbra{\lim_{\eps\to0} \frac{\Phi\pbra{\Phi^{-1}\pbra{\alpha} + \eps} - \Phi\pbra{\Phi^{-1}\pbra{\alpha}}}{\eps}}\\
	&= {\frac 1 {\sqrt{2}}} \rinn\cdot\Phi'\pbra{\Phi^{-1}(\alpha)}\\
	&= {\frac 1 {\sqrt{2}}} \rinn\cdot\varphi\circ\Phi^{-1}(\alpha)
	\end{align*}
	by making the change of variables $\eps := \delta\rinn$ and using the fact that $\varphi = \Phi'$. It follows then from \Cref{prop:gaussian-isop-func-estimate} that 
	\[\TInf[K] \geq {\frac 1 {\sqrt{2}}} \rinn\cdot \pbra{\sqrt{\frac{2}{\pi}}\min(\alpha, 1-\alpha)} \geq {\frac 1 {\sqrt{\pi}}}\Var\sbra{K}\cdot\rinn\] 
	which completes the proof.
\end{proofof}

\medskip

As discussed in Item~1 of \Cref{sec:discussion}, we conjecture that the RHS of \Cref{eq:kkl-convex-sets} can be strengthened to $\Omega\pbra{\Var[K]\log\pbra{\frac{1}{\delta}}}$, which would be the best possible bound by \Cref{eg:solid-cube}.


\section{Towards a Junta Theorem for Convex Sets} 
\label{sec:our-friedgut}

Friedgut's junta theorem \cite{Friedgut:98} is an important result in the analysis of Boolean functions.  It says that Boolean functions with low total influence must be close to juntas; more precisely, if $f: \bn \to \zo$ has $\TInf[f] \leq I$, then $f$ is $\eps$-close to a junta on some set $J$ of at most $2^{O(I/\eps)}$ variables.  Like the KKL theorem, the standard proof of Friedgut's theorem uses the hypercontractive inequality (and is in fact quite similar to the proof of the KKL theorem; see Section~9.6 of \cite{ODbook}).

An easy consequence of Friedgut's junta theorem is that for any low-influence function, averaging out a well-chosen small set of coordinates makes the function have low variance:

\begin{corollary} [Corollary of Friedgut's junta theorem] \label{cor:friedgut}
Let $f: \bn \to \zo$ be a function that has $\TInf[f] \leq I$.  Let $\eps > 0$ and let $f_{-J}: \bn \to [0,1]$ denote the function obtained by ``averaging out'' the coordinates in $J$, i.e. $f_{-J}$ is defined as 
\[
f_{-J}(x^{[n]\setminus J}) := \Ex_{\bx \sim \bits^{J}}[f(\bx^{J},x^{[n]\setminus J})],
\]
where $J$ is the set of $2^{O(I/\eps)}$ variables whose existence is given by Friedgut's junta theorem (so $f_{-J}$ depends only on the coordinates in $[n] \setminus J$).
Then $\Var[f_{-J}] \leq 4\eps.$
\end{corollary}
\begin{proof}
Let $\mu_f = \E[f]=\E[f_{-J}].$ Let $g$ denote the $J$-junta which $\eps$-approximates $f$, and let $\mu_g = \E[g_J]$; note that since $f$ and $g$ are both $0/1$-valued, we have $\E[(f-g)^2] = \eps.$ We have
\begin{align*}
\Var[f_{-J}] &= \E[(f_{-J}-\mu_f)^2]\\
&\leq 2  \left(\E[(f_{-J} -  \mu_g)^2]  + \E[(\mu_f - \mu_g)^2] \right)\\
&\leq 2 \left(\E[(f - g)^2]  + \E[(f -g)^2] \right) = 4\eps. \qedhere
\end{align*}
\end{proof}

In this section we prove a Gaussian space analogue of \Cref{cor:friedgut} for our convex influence notion:

\begin{theorem}[Analogue of \Cref{cor:friedgut} for convex influence] \label{thm:weak-friedgut-convex}
	Let $K\sse\R^n$ be a symmetric convex set with $\TInf[K] \leq I$. For any  $\eps > 0$, 
	there exists a set of orthogonal directions $S = \{v_{i_1}, \ldots, v_{i_\ell}\}$ with $\ell = \exp\pbra{O\pbra{\pbra{I^2/\eps^4}}}$ such that the following holds: For convenience, rename coordinates so that $v_{i_1} = e_1,\dots,v_{i_\ell}=e_\ell$, and define $K_{-S}$ to be the symmetric log-concave function
	\[K_{-S}(x) := \Ex_{(\bx_{1}, \ldots, \bx_{\ell}) \sim {\cal N}(0,1)^\ell}[K(\bx_1,\dots,\bx_\ell,
	x_{\ell+1},\dots,x_n)]\]
	(so $K_{-S}$ depends only on the variables $x_{\ell+1},\dots,x_n$). 
	Then $\Var[K_{-S}] \leq \eps.$
\end{theorem}

\subsection{The Main Technical Ingredient}

The main technical ingredient in our proof of \Cref{thm:weak-friedgut-convex} is the following generalization of \Cref{thm:our-kkl} (our convex influence analogue of the KKL theorem) to symmetric log-concave functions:

\begin{proposition}[KKL for symmetric logconcave functions]
\label{prop:kkl-symmetric-logconcave}
	Let $f: \R^n \to [0,1]$ be a symmetric log-concave function with $\Var[f] \geq \sigma^2$ and $\TInf[f] \leq I$. Then there exists some direction $v^\ast \in \mathbb{S}^{n-1}$ such that 
	\[\Inf_{v^\ast}[f] \geq \Omega\pbra{\sigma^2 e^{-4 \pi I^2/\sigma^4}}.\]
\end{proposition}

\begin{proof}
	For $t\in[0,1]$, define the level set 
	\[A_t := \{x\in\R^n : f(x)\geq t\}.\]
	It is immediate from the log-concavity of $f$ that $A_t$ is a symmetric convex set for all $t \in [0,1]$, and that 
	\[f(x) = \int_{0}^1 A_t(x)\,dt = \Ex_{\bt \in [0,1]}[A_{\bt}(x)],\]
	where we identified $A_t$ with its indicator function. 
	Next, note that for any $x \in \R^n$, by Jensen's inequality we have that 
	\begin{equation} \label{eq:arglebargle}
		\Ex_{\bt}\sbra{\abs{A_{\bt}(x) - \Ex_{\by}\sbra{A_{\bt}(\by)}}^2} \geq \Ex_{\bt}\sbra{\abs{A_{\bt}(x) - \Ex_{\by}\sbra{A_{\bt}(\by)}}}^2 \geq \abs{\Ex_{\bt}\sbra{A_{\bt}(x)} - \Ex_{\bt}\sbra{\Ex_{\by}\sbra{A_{\bt}(\by)}}}^2.
	\end{equation}
	Averaging \Cref{eq:arglebargle} over $\bx \sim {\cal N}(0,1)^n$, the LHS becomes	$\Ex_{\bt}\sbra{\Var\sbra{A_{\bt}}}$ and the RHS becomes $\Var[f]$, so we get that 
	\begin{equation} \label{eq:var-lb}
		\Ex_{\bt}\sbra{\Var\sbra{A_{\bt}}} \geq \Var[f] \geq \sigma^2.
	\end{equation}
	Let $\rinn(A_t)$ denote the in-radius of $A_t$. From the proof of \Cref{thm:our-kkl} (in particular, see \Cref{eq:kkl-first-goal}), we have that 
	\[\TInf[A_t] \geq {\frac 1 {\sqrt{\pi}}} \rinn(A_t)\cdot\Var[A_t] \] 
	and as $\TInf[f] = \int_{0}^1 \TInf[A_t]\,dt$, we have that
	\begin{equation}\label{eq:var-ub}
		I \geq \TInf[f] \geq {\frac 1 {\sqrt{\pi}}} \int_{0}^1\rinn(A_t)\cdot\Var[A_t]\,dt,
		\quad\text{i.e.}\quad
		\sqrt{\pi}I \geq \Ex_{\bt \in [0,1]}[\rinn(A_{\bt}) \Var[A_{\bt}]].
	\end{equation}
	
	Let ${\cal D}$ be the distribution over $[0,1]$ which samples each outcome of $t \in [0,1]$ with probability proportional to $\Var[A_t]$ (so the density function of ${\cal D}$ at $t$ is $\Var[A_t]/\int_0^1 \Var[A_s] ds$). Armed with ${\cal D}$, we may infer from \Cref{eq:var-lb,eq:var-ub} that $
	\Ex_{\bt \sim {\cal D}}[\rinn(A_{\bt})] \leq \sqrt{\pi}I/\sigma^2$, so by Markov's inequality we have that
	\begin{equation} \label{eq:lemon}
	\Pr_{\bt \sim {\cal D}}\left[\rinn(A_{\bt}) \leq 2 \sqrt{\pi}I/\sigma^2\right] \geq 1/2.
	\end{equation}
	
	Recall that $\rinn(A_{t})$ is non-increasing, and let $t^\ast = \inf\{t \in [0,1]: \rinn(A_{t}) \leq 2 \sqrt{\pi}I/\sigma^2\}$.  By definition of ${\cal D}$ and \Cref{eq:lemon}, we have that 
	\begin{equation}
	\label{eq:apple}
	\int_{t = t^\ast}^1 \Var[A_t] dt \geq \sigma^2/2.
	\end{equation} 
	Applying  \Cref{claim:small-inf-big-inradius} to $A_{t^\ast}$, we get that there exists some direction $v^\ast \in \mathbb{S}^{n-1}$ such that   
\begin{equation} \label{eq:nimbu}
\Inf_{v^\ast}[A_{t^\ast}] \geq \frac{\gamma(A_{t^\ast}) e^{-\rinn(A_{t^\ast})^2}}{2^{3/2} \pi} \geq
\frac{e^{-\rinn(A_{t^\ast})^2}}{2^{3/2} \pi} \cdot \Var\sbra{A_{t^\ast}}.
\end{equation}
Since $A_{t} \subseteq A_{t^\ast}$ for $t \geq t^\ast$, it follows from the proof of \Cref{claim:small-inf-big-inradius} (and the fact that $\rinn(A_{t})$ is non-increasing in $t$) that the direction $v^\ast$ has
$\Inf_{v^\ast}[A_{t}] \geq 
\frac{ e^{-\rinn(A_{t^\ast})^2}}{2^{3/2} \pi} \cdot \Var\sbra{A_{t}}$ for all $t \geq t^\ast$.  Hence
\begin{align}
\Inf_{v^\ast}[f] &= \int_{t=0}^1 \Inf_{v^\ast}[A_t] dt \geq
\int_{t=t^\ast}^1 \Inf_{v^\ast}[A_{t}] dt \geq
\frac{ e^{-\rinn(A_{t^\ast})^2}}{2^{3/2} \pi} \cdot \int_{t=t^\ast}^1 \Var\sbra{A_{t}} dt \nonumber \\
&\geq
\frac{ \sigma^2 e^{-\rinn(A_{t^\ast})^2}}{2^{5/2} \pi} \label{eq:banana}\\
&\geq
\frac{ \sigma^2 e^{-4\pi I^2/\sigma^4}}{2^{5/2} \pi} \nonumber
\end{align}
where \Cref{eq:banana} is by \Cref{eq:apple}.
\end{proof}

\subsection{Proof of \Cref{thm:weak-friedgut-convex}}
	If $\Var[K] \leq \epsilon$, then the result clearly holds. If $\Var[K] > \eps$, then by \Cref{prop:kkl-symmetric-logconcave}, there exists some direction $v\in\mathbb{S}^{n-1}$---without loss of generality, say $v = e_1$---such that 
	\[\Inf_{v}[K] = \Inf_{e_1}[K] \geq c\eps\cdot e^{-4 \pi I^2/\eps^2}\]
	for some absolute constant $c > 0$.
Let $K_{-\{e_1\}} : \R^{n-1} \to [0,1]$ be the symmetric log-concave function obtained from $K$ by averaging out the coordinate $e_1$, i.e. we define
	\[K_{-\{e_1\}}(x):= \Ex_{\bx_1 \sim {\cal N}(0,1)}[K(\bx_1,x_2,\dots,x_n)].\]
	It follows from \Cref{obs:influence-averaging} that for all $i \neq 1$, we have $\Inf_{e_i}\sbra{K_{-e_1}} = \Inf_{e_i}[K]$, and so we have 
	\[\TInf\sbra{K_{-e_1}} = \TInf[K] - \Inf_{e_1}[K] \leq I - c\eps\cdot e^{-4 \pi I^2/\eps^2}.\]
	If $\Var[K_{-e_1}] \leq \eps$, then the claimed result holds; if not, then once again by \Cref{prop:kkl-symmetric-logconcave}, there exists some direction---without loss of generality, say $e_2$---such that 
	\[\Inf_{e_2}\sbra{K_{-e_1}} \geq c\eps\cdot e^{-4 \pi I^2/\eps^2},\]
	and we can average out $e_2$ to obtain $K_{-\cbra{e_1, e_2}},$
	\[K_{-\{e_1,e_2\}}(x):= \Ex_{\bx_1,\bx_2 \sim {\cal N}(0,1)}[K(\bx_1,\bx_2,x_3,\dots,x_n)],\]
	with 
	\[\TInf\sbra{K_{-\cbra{e_1, e_2}}} \leq I - 2c\eps\cdot e^{-4 \pi I^2/\eps^2}.\]
	If $\Var\sbra{K_{-\cbra{e_1, e_2}}} \leq \eps$, then the desired result holds; if not, then we repeat as above. Note, however, that the maximum possible number of repetitions (before $\TInf[K_{-S}]$ would become negative, which is impossible) is at most 
	\[\frac{I}{c\eps \cdot e^{-4 \pi I^2/\eps^2}} = \exp\pbra{O( I^2/\eps^2)};\]
	so after at most this many repetitions it must be the case that $\Var\sbra{K_{-S}} \leq \eps.$ This concludes the proof of \Cref{thm:weak-friedgut-convex}. \qed


\section{Sharp Threshold Results for Symmetric Convex Sets\ignore{ under Dilations}}
\label{sec:sharp-threshold}

For any symmetric convex set $K \subseteq \R^n$, we have that $\gamma_\sigma(K) = \gamma(K/\sigma)$, and hence the map $\Psi_K:\sigma  \mapsto \gamma_{\sigma}(K)$ is a non-increasing function of $\sigma$ (since $K/\sigma_1 \subseteq K/\sigma_2$ whenever $\sigma_1 \ge \sigma_2$).
Given this, it is natural to study the rate of decay of $\Psi_K$ for different symmetric convex sets $K \subseteq \R^n$. 

The $S$-inequality (\Cref{prop:s-inequality}) can be interpreted as saying that the slowest rate of decay across all symmetric convex sets of a given volume is achieved by a symmetric strip.
Let $K_{\ast}$ be such a strip, i.e.~we may take $K_{\ast}=\{x \in \mathbb{R}^n: |x_1| \le c_\ast\}$  where $c_\ast = \Theta(\sqrt{\ln(1/\eps)})$ is chosen so that $\Psi_{K_\ast}(1) = 1-\varepsilon$ (and hence $\gamma(K_\ast) = 1-\varepsilon$). With this choice of $c_\ast$, it follows that  $\Psi_{K_\ast}(\sigma) = \varepsilon$ for $\sigma=\tilde{\Theta} (1/\epsilon)$. 
Hence, for the volume of $K_{\ast}$ to shrink from $1-\varepsilon$ to $\varepsilon$, the variance of the underlying Gaussian has to increase very dramatically, by a factor of $\tilde{O} (1/\epsilon^2)$.  Taking, for example, $\varepsilon = 0.01$, we see that in order for the symmetric strip $K_\ast$ to have its Gaussian volume change from $\gamma_{1}(K_\ast)=0.99$ to $\gamma_\sigma(K_\ast)=0.01$, the parameter $\sigma$ must vary over an interval of size $\Theta(1)$, so the strip $K_\ast$ does not exhibit a ``sharp threshold.''

Of course, as we have seen before, the symmetric strip $K_\ast$ has an extremely large (constant) convex influence in the direction $e_1$.  We now show that large individual influences are the only obstacle to sharp thresholds, i.e.~any symmetric convex set in which no direction has large convex influence must exhibit a sharp threshold:

\begin{theorem} [Sharp thresholds for symmetric convex sets with small max influence]~\label{thm:threshold-Friedgut-Kalai}
Let $K \subseteq \mathbb{R}^n$ be a centrally symmetric convex set. Suppose $\varepsilon, \delta>0$ where $\delta  \le \varepsilon^{-10\log(1/\epsilon)}$ 
and $\varepsilon>0$ is sufficiently small (at most some fixed absolute constant).
 Suppose that $\gamma(K) \le 1-\varepsilon$ and $\max_{v \in \mathbb{S}^{n-1}} [\Inf_v(K)] \le \delta$. Then, for $\sigma = 1 + \Theta \pbra{\frac{\ln(1/\varepsilon)}{\sqrt{\ln(\varepsilon/\delta)}}}$, we have $\gamma_{\sigma}(K) \leq\varepsilon$. 
\end{theorem}
Setting $\varepsilon=0.01$ and $\delta =o(1)$, the above theorem implies that for $K$ a  symmetric convex set $K$ with $\max_{v \in \mathbb{S}^{n-1}} [\Inf_v(K)] =o(1)$, it must be the case that $\gamma_{\sigma}(K)$ changes from $0.99$ to $0.01$ as the underlying $\sigma$ changes from $1$ to $1+o(1)$. 

\medskip

\noindent {\bf Discussion.}
\Cref{thm:threshold-Friedgut-Kalai} can be seen as a convex influence analogue of a ``sharp threshold'' result due to Kalai \cite{Kalai:04}. Building on \cite{FriedgutKalai:96}, Kalai \cite{Kalai:04} shows that if $f: \bn \to \zo$ is monotone and its max influence is $o(1)$, then $\mu_p(f)$ must have a sharp threshold (where $\mu_p(f)$ is the expectation of $f$ under the $p$-biased measure) (see also Theorem~3.8 of~\cite{KalaiSafra:05}). This is closely analogous to~\Cref{thm:threshold-Friedgut-Kalai}, which establishes a sharp threshold for $\gamma_{\sigma}(K)$ under the assumption that the max convex influence of $K$ is $o(1)$. We note an interesting quantitative distinction between \Cref{thm:threshold-Friedgut-Kalai} and the result of \cite{Kalai:04}:  if the max influence of a monotone $f: \bn \to \zo$ function is $\delta$, then \cite{Kalai:04} shows that $\mu_p(f)$ goes from $0.01$ to $0.99$ in an interval of width $\approx 1/\mathsf{poly}(\log \log (1/\delta))$ (see the discussion following Theorem~3.8 of \cite{KalaiSafra:05}). In contrast, \Cref{thm:threshold-Friedgut-Kalai} shows that $\gamma_{\sigma}(K)$ goes from $0.01$ to $0.99$ in an interval of width $\approx 1/\sqrt{\log (1/\delta)}$, thus establishing an exponentially ``sharper threshold'' in the convex setting.\footnote{Roughly speaking, the extra exponential factor in \cite{Kalai:04} arises because of Friedgut's junta theorem; our proof takes a different path and does not incur this quantitative factor.}

\medskip

\begin{proofof}{Theorem~\ref{thm:threshold-Friedgut-Kalai}}
We may assume that $\gamma(K) \ge \varepsilon$, since otherwise, there is nothing to prove.
Let $\rinn=\rinn(K)$ be the in-radius of $K$.
By \Cref{claim:small-inf-big-inradius} we get that
\begin{equation}~\label{eq:lb-rinn-K1}
\rinn \ge \sqrt{\ln \bigg( \frac{\gamma(K)}{2^{3/2} \pi \delta}\bigg)} \ge \sqrt{\ln \bigg( \frac{\varepsilon}{2^{3/2} \pi \delta}\bigg)} 
\end{equation} 
(note that our assumptions on $\delta$ and $\varepsilon$ imply that the right hand side of \eqref{eq:lb-rinn-K1} is well-defined). 
Next, we observe that a \emph{mutatis mutandis} modification of the proof of \Cref{eq:kkl-first-goal} gives that
\begin{equation}~\label{eq:influence-lb-sigma}
\TInf^{(\sigma)}[K] \ge \frac{1}{\sqrt{\pi}} \cdot  \rinn \cdot \Var_{\sigma}[K]. 
\end{equation}
We further recall that by our Marguilis-Russo formula for symmetric convex sets (\Cref{prop:russo-margulis-convex}), we have 
\begin{equation}~\label{eq:russo-m-1}
\frac{d \gamma_{\sigma}(K)}{d\sigma^2} = -\frac{1}{\sigma^2 \sqrt{2}} \TInf^{(\sigma)}[K].
\end{equation}
Combining \eqref{eq:lb-rinn-K1}, \eqref{eq:influence-lb-sigma} and \eqref{eq:russo-m-1}, we get  that 
\begin{equation}\nonumber
 \frac{d \gamma_{\sigma}(K)}{d\sigma^2} \leq  -\frac{1}{\sqrt{2 \pi}\sigma^2} \cdot \Var_{\sigma}[K] \cdot \sqrt{\ln \bigg( \frac{\varepsilon}{2^{3/2} \pi \delta}\bigg)}. 
\end{equation}
Expressing $\Var_{\sigma}[K]$ as $\gamma_{\sigma}(K) \cdot (1-\gamma_{\sigma}(K))$ and ``solving'' the above differential equation for $\gamma_{\sigma}(K)$, we get that
\begin{equation} \label{eq:potato}
\ln \bigg( \frac{\gamma_{\sigma}(K)}{1-\gamma_{\sigma}(K) }\bigg) - \ln  \bigg( \frac{\gamma(K)}{1-\gamma(K) }\bigg) \le  \frac{-1}{\sqrt{2 \pi}} \cdot \sqrt{\ln \bigg( \frac{\varepsilon}{2^{3/2} \pi \delta}\bigg)} \cdot 2 \ln \sigma. 
\end{equation}
Using the assumption that $\gamma(K) \leq 1-\epsilon$, it follows that  for $\sigma \ge 1$, we have
\[
\ln \bigg( \frac{\gamma_{\sigma}(K)}{1-\gamma_{\sigma}(K) }\bigg) \le \ln (1/\epsilon) + \frac{-1}{\sqrt{2 \pi}} \cdot \sqrt{\ln \bigg( \frac{\varepsilon}{2^{3/2} \pi \delta}\bigg)} \cdot 2 \ln \sigma. 
\]
Recalling the assumption that $\delta  \le \varepsilon^{-10\log(1/\epsilon)}$ , we see that choosing 
\[
\sigma  = 1 + \Theta \pbra{\frac{\ln(1/\varepsilon)}{\sqrt{\ln(\varepsilon/\delta)}}},
\]
we get $\gamma_{\sigma}(K) \leq \varepsilon$ as claimed.
\end{proofof}

\begin{remark} We close this section by noting that the type of threshold phenomenon studied here has previously been considered in geometric functional analysis. 
 In particular, the seminal work of Milman \cite{Milman:71}, using concentration of measure to establish Dvoretzky's theorem \cite{Dvoretzky:61}
 on almost Euclidean sections of symmetric convex sets, implies a type of threshold phenomenon for symmetric convex sets. Milman's result can be shown to imply that if the ``Dvoretzky number'' of a symmetric convex set is $\omega_n(1)$, then the set must exhibit a type of sharp threshold behavior.  Indeed, Milman's theorem can be used to give an alternate proof of a result that is similar to \Cref{thm:threshold-Friedgut-Kalai}. 
 \end{remark}

\section{A Robust Kruskal-Katona Analogue\ignore{Shell-Density Increment ()} for Symmetric Convex Sets}
\label{sec:robust-kk}

Recall from \Cref{eq:sdf} that for a symmetric convex set $K \subseteq \mathbb{R}^n$, the shell density function $\alpha_K : [0,\infty) \rightarrow [0,1]$ is defined to be
$\alpha_K(r) \coloneqq \Prx_{\bx \in \mathbb{S}^{n-1}_r}[\bx \in K]$, and that $\alpha_K(\cdot)$ is non-increasing. 
In \cite{DS21-weak-learning}, De~and~Servedio established the following \emph{quantitative} lower bound on the rate of decay of $\alpha_K(\cdot)$: 

\begin{theorem} [Theorem~12 of~\cite{DS21-weak-learning}] ~\label{thm:DS21-weak-learning}
Let $K \subseteq \mathbb{R}^n$ be a convex body that has in-radius $\rinn>0$. Then for $r>\rinn$ such that $\min \{\alpha_K(r), (1-\alpha_K(r))\} \ge e^{-n/4}$, as $\Delta r \rightarrow 0^+$ we have that
\[
\alpha_K(r -\Delta r) - \alpha_K(r) \ge \Omega \bigg(\frac{\rinn  \cdot \sqrt{n} \cdot \Delta r}{r^2} \bigg) \alpha_K(r) (1-\alpha_K(r)).
\]
\end{theorem}
As alluded to in Item~1 of \Cref{sec:intro}, the above result can be used to obtain a Kruskal-Katona type theorem for centrally symmetric convex sets. In particular, we have the following corollary:

\begin{corollary}~\label{corr:DS21-weak-learning}
Let $K \subseteq \mathbb{R}^n$ be a  symmetric convex set and $r = \Theta(\sqrt{n})$ be such that $\alpha_K(r) \in [1/10, 9/10]$. Then, as $\varepsilon \rightarrow 0^+$, we have that 
\[
\alpha_K(r(1-\varepsilon)) - \alpha_K(r) = \Omega(\varepsilon).  
\]
\end{corollary}
\begin{proof}
Let $\rinn$ denote the in-radius of $K$, so for any $\zeta>0$, there is a point $z_\ast$ such that $z_\ast \not \in K$ and $\Vert z_\ast \Vert_2 = \rinn + \zeta$. 
By the separating hyperplane theorem, it follows that 
there is a unit vector $\hat{v} \in \mathbb{R}^n$ such that 
\begin{equation}~\label{eq:inclusion-1}
K \subseteq  K_{\ast} \coloneqq \{x \in \mathbb{R}^n: |\hat{v} \cdot x| \le \rinn+ \zeta\}. 
\end{equation}
We next upper bound $\alpha_{K_\ast}(r)$. For this, without loss of generality, we may assume that $\hat{v}=e_1$. We have
\[
\alpha_{K_\ast}(r) = \Prx_{y \in \mathbb{S}^{n-1}_r} [|y_1| \le \rinn+ \zeta] \leq O \bigg(\frac{(\rinn +\zeta) \cdot \sqrt{n}}{r}\bigg),
\]
where the upper bound is an easy consequence of well-known concentration of measure results for the $n$-dimensional unit sphere.
Now, using \eqref{eq:inclusion-1} and letting $\zeta \rightarrow 0$, we have 
\[
\alpha_K(r) \le \alpha_{K_\ast}(r)  \leq O \bigg(\frac{\rinn\cdot \sqrt{n}}{r}\bigg).
\]
Since $\alpha_K(r) \ge 0.1$ by assumption, it follows that $\rinn = \Omega(1)$. \Cref{corr:DS21-weak-learning} now follows from~\Cref{thm:DS21-weak-learning}. 
\end{proof}

\medskip

\noindent {\bf A Robust Analogue of Kruskal-Katona.}  
The lower bound given by \Cref{corr:DS21-weak-learning} cannot be improved in general; for example, the convex set $K= \Dict_{e_1}  \coloneqq \{x: |x_1| \le 1\}$ satisfies the conditions of \Cref{corr:DS21-weak-learning} and has
\[
\alpha_{\Dict_{e_1}}(r(1-\varepsilon)) - \alpha_{\Dict_{e_1}}(r) = \Theta(\varepsilon)
\] 
for $r=\Theta(\sqrt{n}).$  This is closely analogous to how the $\Omega(1/n)$ density increment of the original Kruskal-Katona theorem for monotone Boolean functions (recall Item~1 of \Cref{sec:intro}) cannot be improved in general because of functions like the Boolean dictator function $f(x)=x_1$.  However, if ``large single-coordinate influences'' are disallowed then stronger forms of the Kruskal-Katona theorem, with larger density increments, hold for monotone Boolean functions. In particular, O'Donnell and Wimmer proved the following ``robust'' version of the Kruskal-Katona theorem:
\begin{theorem} [Theorem~1.3 of ~\cite{OWimmer:09}] \label{thm:KK-stability}
Let $f: \{\pm 1\}^n \rightarrow \{0,1\}$ be a monotone function and let $n/10 \le j \le 9n/10.$ 
If $1/10 \leq \mu_j(f) \le 9/10$ and it holds for all $i\in[n]$ that
\begin{equation}~\label{eq:low-influence-cond}
\bigg|\Prx_{\bx \sim \binom{[n]}{j}} [f(\bx) = 1 | \bx_i = 1] -\Prx_{\bx \sim \binom{[n]}{j}} [f(\bx) = 1 | \bx_i = -1] \bigg| \le \frac{1}{n^{1/10}},
\end{equation}
then
$$  \mu_{j+1} (f) \ge \mu_j(f)  + \Omega \bigg( \frac{\log n}{n} \bigg).$$
\end{theorem}
In words, under condition \Cref{eq:low-influence-cond} (which is akin to saying that each variable $x_i$ has ``low influence on $f$''), the much larger density increment $\Omega(\log(n)/n)$ must hold.

Using our notion of convex influences, we now establish a robust version of \Cref{corr:DS21-weak-learning} which is similar in spirit to the Boolean ``robust Kruskal-Katona'' result given by \Cref{thm:KK-stability}.  Intuitively, our result says that if all convex influences are small, then we get a stronger density increment than \Cref{corr:DS21-weak-learning}:

\ignore{

}

\begin{theorem}~\label{thm:KK-robust-convex}
Let $K \subseteq \mathbb{R}^n$ be a centrally symmetric convex set and $ \sqrt{n} \le r = \Theta( \sqrt{n})$ be such that 
$\alpha_K(r) \in [1/10, 9/10]$. If $\Inf_{v}[K] \le \delta$ for all $v \in \mathbb{S}^{n-1}$ then as $\varepsilon \rightarrow 0^+$ we have that
\[
\alpha_K(r(1-\varepsilon)) - \alpha_K(r) = \Omega(\varepsilon \sqrt{\ln(1/\delta)}). 
\]
\end{theorem}

\begin{proofof}{Theorem~\ref{thm:KK-robust-convex}}
We begin by proving that $\gamma(K) = \Theta(1)$. Note that 
\begin{eqnarray*}
\gamma(K) = \int_{r=0}^\infty \alpha_K(r) \cdot \chi_n(r) dr\ge \int_{r=0}^{\sqrt{n}} \alpha_K(r) \cdot  \chi_n(r) dr, 
\end{eqnarray*}
where $\chi_n(\cdot)$ is the pdf of the $\chi$-distribution with $n$-degrees of freedom. Now, since $\alpha_K(\cdot)$ is non-increasing and $\int_{r=0}^{\sqrt{n}} \chi_n(r) = \Omega(1)$, it must be the case that
\begin{eqnarray}~\label{eq:lb-volume-K}
\gamma(K) \ge  \alpha_K(\sqrt{n}) \cdot \int_{r=0}^{\sqrt{n}}  \chi_n(r) dr = \Theta(1),
\end{eqnarray}
where the last equality uses the fact that $r \geq \sqrt{n}$ and $\alpha_K(r) \ge 1/10$. 

Let $\rinn$ denote the in-radius of $K$. Exactly as reasoned in the proof of~\Cref{corr:DS21-weak-learning}, there exists a unit vector $v \in \mathbb{S}^{n-1}$ such that $K \subseteq \{x \in \mathbb{R}^n: |v \cdot x| \le \rinn + \zeta\}$ for any $\zeta>0$. Since $\gamma(K)=\Omega(1)$, it now follows from \Cref{claim:small-inf-large-slab} that there is a direction $v$ such that 
\[
\Inf_v[K]  = \Omega (e^{-(\rinn + \zeta)^2}). 
\]
By our hypothesis, we have that $\Inf_v[K] \le \delta$, so taking $\zeta \rightarrow 0$ we get that
$\rinn = \Omega(\sqrt{\ln (1/\delta)})$ (note that we may assume $\delta$ is at most some sufficiently small constant, since otherwise the claimed result is given by~\Cref{corr:DS21-weak-learning}).
 We now apply~\Cref{thm:DS21-weak-learning} to obtain that 
\[
\alpha_K(r(1-\varepsilon)) - \alpha_K(r) = \Omega(\varepsilon \sqrt{\ln(1/\delta)}), 
\]
thus proving \Cref{thm:KK-robust-convex}. 
\end{proofof}






\section*{Acknowledgements} 
A.D.~is supported by NSF grants CCF 1910534 and CCF 1926872.  S.N.~is supported
by NSF grants CCF-1563155 and by 
CCF-1763970.  R.A.S.~is supported by NSF grants CCF-1814873, IIS-1838154,
CCF-1563155, and by the Simons Collaboration on Algorithms and Geometry.  This
material is based upon work supported by the National Science Foundation under
grant numbers listed above. Any opinions, findings and conclusions or
recommendations expressed in this material are those of the authors and do not
necessarily reflect the views of the National Science Foundation (NSF). This work was done while A.D. was participating in the ``Probability, Geometry, and Computation in High Dimensions'' program at the Simons
Institute for the Theory of Computing.

\bibliography{allrefs}{}
\bibliographystyle{alpha}

\appendix

\section{Omitted Proofs from \Cref{sec:influence-basics}} \label{appendix:sec-3}

\subsection{Proof of \Cref{prop:influence-nonneg}}

We will require the following fact about log-concave functions: 

\begin{fact} [Lemma~4.7 of \cite{Vempalafocs10}]
\label{fact:1dlogconcave}
Let $g: \R \to \R^+$ be a  log-concave function such that \[\Ex_{\bx \sim \calN(0,1)}[\bx g(\bx)]=0.\] Then $\E[ \bx^2 g(\bx)] \le \E[g(\bx)],$ with equality if and only if $g$ is a constant function.
\end{fact}

We will also require the following Brunn-Minkowski-type inequality over Gaussian space, as well as a recent characterization of the equality case:

\begin{proposition}[Ehrhard-Borell inequality, \cite{Ehrhard:83,Borell:03,Borell:08}] 
\label{prop:ehrhard-borell}
	Let $A, B \subseteq \R^n$ be Borel sets, identified with their indicator functions. Then 
	\begin{equation} \label{eq:ehrhard}
		\Phi^{-1}\pbra{\gamma_n\pbra{\lambda A + (1-\lambda)B}} \geq \lambda\Phi^{-1}\pbra{\gamma_n\pbra{A}} + (1-\lambda)\Phi^{-1}\pbra{\gamma_n\pbra{B}} 
	\end{equation}
	where\ignore{$\Phi : \R \to [0,1]$ denotes the cumulative distribution function of the standard, one-dimensional Gaussian distribution, $\gamma_n$ denotes the $n$-dimensional standard Gaussian measure, and} $\lambda A +(1-\lambda)B := \{ \lambda x+ (1-\lambda)y : x\in A, y\in B \}$ is the \emph{Minkowski sum} of $\lambda A$ and $(1-\lambda)B$.
\end{proposition}

\begin{proposition}[Theorem 1.2 of \cite{rvh-equality}] \label{fact:rvh-equality-ehrhard}
Equality holds in the Ehrhard-Borell inequality (\Cref{eq:ehrhard}) if and only if either
	\begin{itemize}
		\item $A$ and $B$ are parallel halfspaces, i.e. we have
		\[A = \{ x : \abra{a, x} + b_1 \geq 0 \} \qquad\text{and}\qquad B = \{ x : \abra{a, x} + b_2 \geq 0 \}\] 
		for some $a \in \R^n$, and $b_1, b_2 \in \R$; or

		\item $A$ and $B$ are convex sets with $A = B$.
	\end{itemize}
\end{proposition}

\begin{proof}[Proof of \Cref{prop:influence-nonneg}]

	Without loss of generality, let $v = e_1$. We have
	\begin{align}
		\Inf_{e_1}[K] = -\wt{K}(2e_1) &= \Ex_{\bx\sim\calN(0,1)^n}\sbra{-K(\bx)h_2(\bx_1)}\nonumber\\
		&= \Ex_{\bx_1\sim\calN(0,1)}\sbra{-\pbra{\underbrace{\Ex_{(\bx_2, \ldots, \bx_n)\sim\calN(0,1)^{n-1}}\sbra{K(\bx_1, \ldots, \bx_n)}}_{=: K_{e_1}(\bx_1)}}h_2(\bx_1)} \label{num:cucumber}\\
		& = -\wt{K_{e_1}}(2), \nonumber
	\end{align}
	where $g$ is a univariate function. From \Cref{fact:marginal-log-concave}, it follows that $K_{e_1}$ is log-concave, and $K_{e_1}$ is symmetric since $K$ is symmetric, so $\E_{\bx_1 \sim \calN(0,1)}[\bx_1 g(\bx_1)]=0$. Hence, using the fact that $h_2(x_1) = (x_1^2-1)/\sqrt{2}$, we get that
	\[
	\Inf_{e_1}[K] = -\Ex_{\bx_1 \sim \calN(0,1)}[h_2(\bx_1) K_{e_1}(\bx_1)] = {\frac 1 {\sqrt{2}}} \cdot \Ex_{\bx_1 \sim \calN(0,1)}\big[K_{e_1}(\bx_1)(1-\bx_1^2)  \big] \ge 0, 
	\]
where the inequality is by~\Cref{fact:1dlogconcave}.	

Next, we move to  the characterization of $\Inf_{e_1}[K]=0$. Note that if $K(x) = K(y)$ whenever $x_{e_1^\perp} = y_{e_1^\perp}$ (i.e. $K(x) = K(y)$ whenever $(x_2, \ldots, x_n)= (y_2, \ldots, y_2)$), then the function $K$ does not depend on the variable $x_1$. This lets us re-express \Cref{num:cucumber} as 
	\[
	\Ex_{\bx_1\sim\calN(0,1)}[-h_2(\bx_1)] \Ex_{(\bx_2, \ldots, \bx_n)\sim\calN(0,1)^{n-1}}[K(\cdot, \bx_2, \ldots, \bx_n)]. 
	\]
	As the first term in the above product is zero, we conclude that $\Inf_{e_1}[K]=0$. 
	
	To see the reverse direction, suppose $\Inf_{e_1}[K]=0$. From \Cref{num:cucumber} and~\Cref{fact:1dlogconcave}, it follows that $K_{e_1}(\cdot)$ is a constant function. Now, for any $\alpha \in \mathbb{R}$, define $K_{\alpha} \subset \R^{n-1}$ as follows: \[
	K_\alpha :=\{ (x_2, \ldots, x_n) : (\alpha, x_2, \ldots, x_n) \in K\}.\] 
	Thus, $K_{\alpha}$ is the convex set obtained by intersecting $K$ with the affine plane $\{x \in \R^n: x_1=\alpha\}$. 
Observe that $K_{e_1}(\alpha)$ is the $(n-1)$-dimensional Gaussian measure of $K_{\alpha}$. Let $K_\alpha^* := {{\frac 1 2}}(K_{\alpha} + K_{-\alpha})$. Note that $K_\alpha^* \subseteq \R^{n-1}$ is a centrally symmetric, convex set, and that $K_\alpha^* \subseteq K_0$ because of convexity. By the Ehrhard-Borell inequality, we have 
\[ 
\Phi^{-1}\pbra{\gamma_{n-1}\pbra{\frac{K_\alpha + K_{-\alpha}}{2}}} \geq \frac{1}{2}\Phi^{-1}\pbra{\gamma_{n-1}(K_\alpha)} + \frac{1}{2}\Phi^{-1}\pbra{\gamma_{n-1}(K_{-\alpha)}}.
\]
However, $\gamma_{n-1}(K_\alpha) = \gamma_{n-1}(K_{-\alpha})$ because $K$ is centrally symmetric, so it follows that $\gamma_{n-1}(K_\alpha^*) \geq \gamma_{n-1}(K_\alpha)$. From our earlier observation that $K_\alpha^* \subseteq K_0$ and that $K_{e_1}(\cdot)$ is constant (which implies $\gamma_{n-1}(K_\alpha) = \gamma_{n-1}(K_0)$), it follows that $K_0 = K_\alpha^*$ up to a set of measure zero. In other words, we have equality in the application of the Ehrhard-Borell inequality above, and so by \Cref{fact:rvh-equality-ehrhard}, we must have $K_\alpha = K_{-\alpha}$ (since $K_\alpha$ and $K_{-\alpha}$ cannot be parallel halfspaces). Consequently, {up to a set of measure zero, we have that}
\[K_0 = K_\alpha^* = \frac{K_\alpha + K_{-\alpha}}{2} = \frac{K_\alpha + K_{\alpha}}{2} = K_\alpha.\] 
As this is true for all $\alpha \in \R$, it follows that {up to a set of measure zero,} $K(x) = K(y)$ if $(x_2, \ldots, x_n) = (y_2, \ldots, y_n)$ (where we used the fact that $K(x) = K_{x_1}(x_2, \ldots, x_n)$). 
\end{proof}

\subsection{Proof of \Cref{claim:small-inf-big-inradius}}

We will use the following Brascamp--Lieb-type inequality.

\begin{lemma} [Final assertion of Lemma~4.7 of \cite{Vempalafocs10}] \label{lem:vempala}
If $g: \R \to \R_{\geq 0}$ is log-concave and symmetric and supported in $[-c,c]$, then
\[
{\frac {\int_{-c}^c x^2 e^{-x^2 / 2} g(x) dx}{\int_{-c}^c e^{-x^2 / 2} g(x) dx}} \leq 1 - {\frac 1 {2 \pi}} e^{-c^2}.
\]
\end{lemma}

We use this in the proof of the following claim, which will easily yield \Cref{claim:small-inf-big-inradius}:

\begin{proposition} \label{claim:small-inf-large-slab}
Let $K \subseteq \R^n$ be a centrally symmetric convex set with $\gamma(K) \geq \Delta$, and let $v \in \mathbb{S}^{n-1}$ be a unit vector such that $K \subseteq \{x \in \R^n: |v \cdot x| \leq c\}$. Then we have
\[\Inf_v[K] \geq \frac{\Delta e^{-c^2}}{2^{3/2} \pi}.\]
\end{proposition}

\begin{proof}[Proof of \Cref{claim:small-inf-large-slab}]
For ease of notation, we take $v=e_1$ and so $K \subseteq \{x \in \R^n: |x_1| \leq c\}.$ From \Cref{eq:inf-averaging-no-change,eq:inf-averaging-def}, we have that
\begin{equation} \label{eq:applesauce}
	\Inf_v[K] = \Inf_{e_1}[K] = \TInf\sbra{K_{e_1}} = {\frac 1 {2 \sqrt{\pi}}} \int_\R K_{e_i}(x)(1-x^2) e^{-x^2/2}\,dx
\end{equation}
where $K_{e_1} : \R \to [0,1]$ is the symmetric log-concave function given by 
\[K_{e_1}(x) := \Ex_{\bx\sim\calN(0,1)^{n-1}}\sbra{K(x, \bx_2, \ldots,  \bx_n)}.\]
As $K(x)=0$ when $|x_1|>c$ we have that  $\supp(g) \sse [-c, c]$ and so it follows from \Cref{eq:applesauce} that
\begin{align}
\Inf_v[K] = {\frac 1 {2 \sqrt{\pi}}} \int_{-c}^c K_{e_1}(x)(1-x^2) e^{-x^2/2}\, dx \label{eq:orange}.
\end{align}  
It follows then from \Cref{lem:vempala} that
\[
\Inf_v[K] \geq {\frac 1 {2^{3/2} \pi}} \pbra{\frac{e^{-c^2}}{\sqrt{2 \pi}}  \int_{-c}^c K_{e_1}(x) e^{-x^2 / 2}\, dx}
=
\frac{\Delta e^{-c^2}}{2^{3/2} \pi}
\]
which completes the proof of \Cref{claim:small-inf-large-slab}.
\end{proof}

\begin{proof}[Proof of \Cref{claim:small-inf-big-inradius}]
By definition of the in-radius and the supporting hyperplane theorem, there must exist
some unit vector $\hat{v} \in \mathbb{R}^n$ such that 
\begin{equation}~\label{eq:inclusion-0}
K \subseteq  K_{\ast} \coloneqq \{x \in \mathbb{R}^n: |\hat{v} \cdot x| \le \rinn\}, \nonumber
\end{equation}
and hence by \Cref{claim:small-inf-large-slab} we get that
\[
\Inf_{\hat{v}}[K] \geq {\frac {\gamma(K) e^{-\rinn^2}}{2^{3/2} \pi}}
\geq {\frac {\Var[K] e^{-\rinn^2}}{2^{3/2} \pi}},
\]
giving \Cref{claim:small-inf-big-inradius} as claimed.
\ignore{
By definition of the in-radius, for any $\zeta>0$, there is a point $z_\ast$ such that $z_\ast \not \in K$ and $\Vert z_\ast \Vert_2 = \rinn + \zeta$. 
By the separating hyperplane theorem, there must exist
some unit vector $\hat{v} \in \mathbb{R}^n$ such that 
\begin{equation}~\label{eq:inclusion-0}
K \subseteq  K_{\ast} \coloneqq \{x \in \mathbb{R}^n: |\hat{v} \cdot x| \le \rinn+ \zeta\}, \nonumber
\end{equation}
and hence by \Cref{claim:small-inf-large-slab} we get that
\[
\Inf_{\hat{v}}[K] \geq {\frac {\gamma(K) e^{-(\rinn + \zeta)^2}}{2^{3/2} \pi}}
\geq {\frac {\Var[K] e^{-(\rinn + \zeta)^2}}{2^{3/2} \pi}}.
\]
Letting $\zeta \to 0^+$, \Cref{claim:small-inf-big-inradius} is proved.
}
\end{proof}
	
\subsection{Proof of \Cref{prop:russo-margulis-convex}}

This is an elementary calculation using the chain rule and the Leibniz rule. 

\begin{proof}[Proof of \Cref{prop:russo-margulis-convex}]
	By the chain rule, we have
	\begin{align*}
		\frac{d}{d\sigma^2}\Ex_{\bx\sim\calN(0,\sigma^2)^n}\sbra{K(\bx)} &= \frac{d}{d\sigma^2}\pbra{\frac{1}{\sqrt{(2\pi\sigma^2)^n}}\underbrace{\int_{K}\exp\pbra{-\frac{\|x\|^2}{2\sigma^2}}\,dx}_{=: A(\sigma^2)}}\\
		&= A(\sigma^2)\frac{d}{d\sigma^2}\pbra{\frac{1}{\sqrt{(2\pi\sigma^2)^n}}} + \frac{1}{\sqrt{(2\pi\sigma^2)^n}}\pbra{\frac{d}{d\sigma^2}A(\sigma^2)}\\
		&= \frac{-nA(\sigma^2)}{2\sigma^2\sqrt{(2\pi \sigma^2)^n}} + \frac{1}{\sqrt{(2\pi\sigma^2)^n}}\pbra{\frac{d}{d\sigma^2}A(\sigma^2)}\\
		&= -\frac{n\Ex_{\bx\sim\calN(0,\sigma^2)^n}\sbra{K(\bx)}}{2\sigma^2} + \frac{1}{\sqrt{(2\pi\sigma^2)^n}}\pbra{\frac{d}{d\sigma^2}A(\sigma^2)}.
	\end{align*}
	Now, using the dominated convergence theorem to commute integration and differentiation, we get
	\begin{align*}
		\frac{1}{\sqrt{(2\pi\sigma^2)^n}}\pbra{\frac{d}{d\sigma^2}A(\sigma^2)} &= \frac{1}{\sqrt{(2\pi\sigma^2)^n}}\pbra{\frac{d}{d\sigma^2}\int_{K}\exp\pbra{-\frac{\|x\|^2}{2\sigma^2}}\,dx}\\
		&= \frac{1}{\sqrt{(2\pi\sigma^2)^n}}\int_{K}\frac{d}{d\sigma^2}\exp\pbra{-\frac{\|x\|^2}{2\sigma^2}}\,dx\\
		&= \frac{1}{\sqrt{(2\pi\sigma^2)^n}}\int_{K}\exp\pbra{-\frac{\|x\|^2}{2\sigma^2}}\frac{\|x\|^2}{2\sigma^4}\,dx\\
		&= \frac{\Ex_{\bx\sim\calN(0,\sigma^2)^n}\sbra{K(\bx)\pbra{\frac{\|\bx\|^2}{\sigma^2}}}}{2\sigma^2}.
	\end{align*}
	This in turn implies that
	\begin{align*}
		\frac{d}{d\sigma^2}\Ex_{\bx\sim\calN(0,\sigma^2)^n}\sbra{K(\bx)} &= -\frac{n\Ex_{\bx\sim\calN(0,\sigma^2)^n}\sbra{K(\bx)}}{2\sigma^2} + \frac{\Ex_{\bx\sim\calN(0,\sigma^2)^n}\sbra{K(\bx)\pbra{\frac{\|\bx\|^2}{\sigma^2}}}}{2\sigma^2}\\
		&= \frac{1}{\sigma^2\sqrt{2}}\Ex_{\bx\sim\calN(0,\sigma^2)^n}\sbra{K(\bx)\pbra{\frac{\pbra{\frac{\|\bx\|^2}{\sigma^2}} - n}{\sqrt{2}}}}\\
		&= \frac{1}{\sigma^2\sqrt{2}}\sum_{i=1}^n\Ex_{\bx\sim\calN(0,\sigma^2)^n} \sbra{K(\bx)\pbra{\frac{\pbra{\frac{\bx_i}{\sigma}}^2 - 1}{\sqrt{2}}}}\\
		&= \frac{1}{\sigma^2\sqrt{2}}\sum_{i=1}^n \wt{K}_\sigma(2e_i)
	\end{align*}
	where we used the fact that $h_2(x) = \frac{x^2 - 1}{\sqrt{2}}$. 
\end{proof}

\end{document}